\documentclass[%
 reprint,
nofootinbib,
amsmath,amssymb,
aip,
physics,
floatfix,
]{revtex4-2}
\usepackage{graphicx}
\usepackage{dcolumn}

\usepackage[mathlines]{lineno}

\usepackage{subcaption}
\usepackage{xfrac}
\usepackage{amsmath}
\usepackage{amsthm}

\def\8{\infty}
\def\oh{\sfrac{1}{2}}

\newcommand*{\I}{\imath}%

\def\const{\textit{ const }}
\def\undertext#1{\vtop{\hbox{#1}\kern 1pt \hrule}}

\def\abs#1{\left| #1\right|}

\def\pd#1{\partial_{#1}}
\def\VEV#1{\left\langle #1\right\rangle}

\def\pbyp#1#2{\frac{\partial#1}{\partial#2}}
\def\ff#1{\frac{\delta}{\delta#1}}

\def\bea{\begin{eqnarray} && &&}
\def\eea{\end{eqnarray}}

\let\oldexp\exp
\renewcommand{\exp}[1]{\oldexp\left(#1\right)}

\def \Schr{Schr\"odinger}

\def\BS{Biot-Savart}
\def\LI {Liouville}
\def\BL {Beltrami}

\def\Pr{\mathit{Pr}}

\newcommand{\Mod}[1]{\ (\mathrm{mod}\ #1)}
\newcommand{\vect}[1]{\boldsymbol{#1}}

\def\XXint#1#2#3{{\setbox0=\hbox{$#1{#2#3}{\int}$}
     \vcenter{\hbox{$#2#3$}}\kern-.5\wd0}}

\renewcommand{\Im}{\textbf{Im }}

\newcommand{\tpmod}[1]{{\@displayfalse\Mod{#1}}}
\usepackage[main=english]{babel}
\usepackage{hyperref}
\usepackage{url}
\hypersetup{colorlinks,allcolors=black}
\usepackage{mathtools}

\newcommand{\pctPDF}[2]{
\begin{figure}
    \centering
    \includegraphics[width=0.45\textwidth]{#1.jpg}
    \caption{#2}
    \label{fig::#1}
\end{figure}
}

\makeatletter
\def\@email#1#2{%
 \endgroup
 \patchcmd{\titleblock@produce}
  {\frontmatter@RRAPformat}
  {\frontmatter@RRAPformat{\produce@RRAP{*#1\href{mailto:#2}{#2}}}\frontmatter@RRAPformat}
  {}{}
}%
\usepackage{lmodern}

\begin{document}

\preprint{APS/123-QED}

\title{Dual Theory of MHD Turbulence}

\author{Alexander Migdal}
\affiliation{Institute for Advanced Study, Princeton, NJ, USA}
\email{amigdal@ias.edu}

\date{\today}
\begin{abstract}
We present an exact analytic solution for decaying incompressible magnetohydrodynamic (MHD) turbulence. Our solution reveals a dual formulation in terms of two interacting Euler ensembles - one for hydrodynamic and another for magnetic circulation. Each ensemble reduces to a pair of quantum pendulums in a Gibbs state. This replaces empirical scaling laws with universal spectra given by quadratures: each spectrum has a single growing scale and a power tail above a threshold wavenumber, $k^{-7/2}$ for the kinetic energy and $k^{-3/2}$ for the magnetic one, and the kinetic energy decays with a slowly drifting index instead of a fixed exponent. A key result of our analysis is the explicit dependence of the solution on the Prandtl number $\Pr = \nu/\eta$. The vorticity and the magnetic field have the same universal correlation function, with the length scales $L$ and $L/\sqrt{\Pr}$. The magnetic force vanishes on this solution: the magnetic field is force-free and aligned with the velocity field. Therefore, the kinetic spectrum is that of hydrodynamic decaying turbulence, and the magnetic spectrum repeats the enstrophy spectrum at the wavenumbers larger by the factor $\sqrt{\Pr}$. This dependence is continuous, with no transition at $\Pr = 1$. We relate our theoretical predictions to recent direct numerical simulations (DNS) and discuss their implications for astrophysical plasmas, fusion devices, and laboratory MHD experiments. Our results provide a  mathematical framework for understanding MHD turbulence and its dependence on fundamental parameters, offering a new perspective on turbulence in highly conducting fluids.
\end{abstract}


\maketitle
\tableofcontents
\section{Introduction}

Magnetohydrodynamics (MHD), the study of electrically conducting fluids in the presence of magnetic fields, is a cornerstone of modern theoretical physics. By coupling the Navier-Stokes equations to Maxwell's equations, MHD provides a framework to describe a wide range of systems, from astrophysical plasmas to laboratory experiments involving fusion devices and liquid metals. This section outlines the role of MHD in astrophysics and plasma physics, and provides a brief overview of the state of the art in MHD theory and numerical simulations.

\subsection{Role of MHD in Astrophysics}

In astrophysics, MHD is essential for understanding the dynamics of plasmas in stars, galaxies, and the interstellar medium (ISM). Magnetic fields are deeply involved in many fundamental processes, including:
\begin{itemize}
    \item \textbf{Star Formation:} Magnetic fields influence the collapse of molecular clouds and regulate the fragmentation process, thereby playing a key role in star formation \cite{McKee2007,Crutcher2012}.
    \item \textbf{Accretion Disks and Jets:} MHD governs the behavior of accretion disks surrounding black holes and protostars, as well as the launching and collimation of astrophysical jets \cite{Balbus1991,Blandford1982}.
    \item \textbf{Galactic and Interstellar Processes:} In the ISM and galactic halos, MHD turbulence mediates energy dissipation, cosmic ray propagation, and the amplification of magnetic fields via dynamo mechanisms \cite{Schekochihin2004,Federrath2016}.
    \item \textbf{Solar and Stellar Phenomena:} The heating of the solar corona, solar wind acceleration, and magnetic reconnection events, such as solar flares, are all MHD-driven phenomena \cite{Parker1958,Aschwanden2004}.
\end{itemize}
On larger scales, MHD effects are critical to understanding the evolution of cosmic magnetic fields, from galaxy clusters to the cosmic web \cite{Ryu2009,Subramanian2016}, shedding light on the interplay between turbulence, reconnection, and magnetic field amplification in these vast structures.

\subsection{Role of MHD in Plasma Physics}

In plasma physics, MHD provides the theoretical framework for describing magnetically confined fusion plasmas, such as those in tokamaks and stellarators. It governs:
\begin{itemize}
    \item \textbf{Stability Analysis:} MHD captures instabilities such as tearing modes, kink instabilities, and ballooning modes, which are critical for understanding confinement and energy transport \cite{Furth1963,Freidberg2014}.
    \item \textbf{Turbulence and Dissipation:} MHD turbulence mediates energy dissipation and reconnection processes, enabling small-scale energy transfer in both fusion plasmas and laboratory experiments \cite{Chandran2004,Zweibel2009}.
    \item \textbf{Magnetic Confinement:} MHD governs the dynamics of magnetic fields used to confine high-temperature plasmas, a critical element of controlled nuclear fusion research \cite{Wagner2013,Hazeltine2013}.
\end{itemize}
MHD also has applications in liquid metal systems, geophysical fluid dynamics, and engineering applications where magnetic fields interact with conducting fluids \cite{Davidson2001}.

\subsection{State of the Art in MHD Theory}

The theoretical understanding of MHD spans both laminar and turbulent regimes. MHD turbulence, in particular, has been the focus of significant research, as it governs energy transfer in magnetized plasmas. Key developments in MHD theory include:
\begin{itemize}
    \item \textbf{Energy Cascades:} The Kolmogorov-like cascade in MHD turbulence is modified by the presence of magnetic fields, leading to anisotropic energy spectra and distinct scaling regimes, such as Alfvénic turbulence \cite{Goldreich1995,Biskamp2003}.
    \item \textbf{Reduced MHD Models:} In strongly magnetized systems, reduced MHD models simplify the equations of motion while capturing the essential physics, making them valuable for analytical and numerical studies \cite{Strauss1976,Montgomery1983}.
    \item \textbf{Dynamo Theory:} Dynamo mechanisms describe how turbulence can amplify magnetic fields, a key process in astrophysical systems like stars and galaxies \cite{Moffatt1979,Brandenburg2005}.
    \item \textbf{Phenomenological Models:} Approaches such as the Iroshnikov-Kraichnan model and Goldreich-Sridhar theory provide scaling predictions for MHD turbulence, although their validity remains a topic of debate \cite{IroshnikovKraichnan2013}.
\end{itemize}
Despite these advances, exact solutions to MHD turbulence remain elusive, as the highly nonlinear nature of the governing equations presents a formidable challenge.

\subsection{State of the Art in Numerical Simulations}

Numerical simulations have become an indispensable tool for studying MHD, particularly in the turbulent regime. High-resolution direct numerical simulations (DNS) have provided valuable insights into the structure and dynamics of MHD turbulence, including:
\begin{itemize}
    \item \textbf{Energy Cascades and Spectra:} Simulations have (approximately) confirmed scaling laws for anisotropic turbulence and revealed new phenomena, such as reconnection-driven turbulence \cite{Mason2009,Cho2009}.
    \item \textbf{Reconnection and Dissipation:} Simulations capture the role of magnetic reconnection as a mechanism for energy dissipation in turbulent systems, particularly in astrophysical plasmas \cite{Daughton2011,Loureiro2017}.
    \item \textbf{Astrophysical Applications:} MHD simulations are used to model accretion disks, jets, galactic dynamos, and ISM turbulence, helping to bridge theory with observations \cite{Hawley1995,Federrath2021}.
    \item \textbf{Fusion Applications:} In the context of fusion, simulations inform the design of tokamaks and stellarators by predicting plasma behavior and assessing stability \cite{Jardin2012}.
\end{itemize}
Despite the progress, numerical simulations face limitations due to computational constraints. Achieving the high Reynolds and magnetic Reynolds numbers characteristic of astrophysical plasmas remains challenging. Furthermore, capturing kinetic effects beyond the fluid approximation, such as collisionless reconnection and particle acceleration, often requires hybrid or fully kinetic simulations, which are computationally expensive \cite{Birn2001,Hesse2011}.

The situation is clearer in the purely hydrodynamic case. A
comprehensive long-time DNS campaign \cite{SreeniAkash2025} showed
that the energy-decay exponent depends on the infrared part of the
initial spectrum, while the bulk of the spectrum, which dominates the
enstrophy, does not. The loop-space theory in its current form
\cite{MigdalQPTF2026} has both features. The Euler ensemble
reduces to a pair of quantum pendulums; its energy spectrum has a fixed
shape with a single growing scale $L(t)\propto\sqrt{t}$; and the
energy, an integral dominated by the lowest wavenumbers of the flow,
decays with an index that depends on the infrared cutoff instead of
being a universal number. This solution supersedes the saddle-point
(instanton) evaluation of the same ensemble
\cite{migdal2024quantum}; its comparison with the DNS and with wind-tunnel data is
the subject of a separate paper (A.~Migdal and W.~Wang, in preparation). This
precedent directly motivates the present MHD analysis.

\section{Motivation and Scope of This Work}

Building on recent advances in the analytical framework of loop equations \cite{migdal2023exact, migdal2024quantum}, this work reduces the problem of decaying magnetohydrodynamic (MHD) turbulence to a dual system of random walks on regular star polygons. This novel formulation reveals an unexpected connection to string theory, where the target space is discrete rather than continuous. Unlike phenomenological models, our approach provides a mathematically justified duality, akin in spirit to AdS/CFT, without resorting to approximations or ad hoc closures.

In this sense, we treat MHD turbulence as a problem in mathematical physics—solved from first principles—rather than as a phenomenological modeling exercise or a purely numerical simulation.

In the turbulent limit—defined by an infinite Reynolds number at arbitrary Prandtl number—the correlation functions of vorticity and magnetic fields, along with macroscopic observables such as the energy spectrum and decay rate, emerge naturally from this discrete string-theoretic formulation. The theory remains solvable in the quasiclassical (turbulent) limit, allowing for explicit expressions for the decaying energy spectrum in quadrature.

This framework not only generalizes prior results from pure hydrodynamic turbulence but also introduces new structural insights into MHD dynamics. In particular, the explicit dependence on the Prandtl number provides a well-defined geometric structure, positioning the problem at the intersection of turbulence theory, discrete geometry, and 
theory of integrable systems.

The most striking prediction is the relation between the two fields. The magnetic field has the same universal correlation function as the vorticity, with the length scale reduced by the factor $\sqrt{\Pr}$, while the kinetic spectrum does not depend on the Prandtl number: the magnetic force vanishes on our solution, where the magnetic field is force-free and aligned with the velocity field. Both spectra and their decay are now described by a quantitative microscopic theory.
\section{MHD equations for two circulations}
The loop functional is defined as a phase factor associated with velocity and vector potential circulations, averaged over the ensemble of the solutions of the MHD equations
\begin{eqnarray}
  &&  \Psi_v[t,C] = \VEV{\exp{\frac{\I (\Gamma_v[C] )  }{\nu}}}_{sol};\\
  && \Gamma_v[C] = \oint_C \vect v(\vect r) \cdot d \vect r;\\
  && \Psi_a[t,C] = \VEV{\exp{\frac{\I (\Gamma_a[C] )  }{\eta}}}_{sol};\\
  && \Gamma_a[C] = \oint_C \vect a(\vect r) \cdot d \vect r;
\end{eqnarray}
We use viscosity $\nu$ and ``magnetic viscosity'' $\eta$ as units of each circulation. Both have the same dimension $L^2/T$ as the Planck's constant $\hbar$.
The viscosities will play the same role in our theory as Planck's constant in quantum mechanics.
We include some factors in normalizing the vector potential $\vect a(\vect r)$ to make it the same dimension $L/T$ as the velocity field, so that each of the two circulations is measured in units of its viscosity. The magnetic field $\vect b = \vect \nabla \times \vect a$ then has the dimension $1/T$ of the vorticity. The magnetic field of the same dimension $L/T$ as the velocity (so-called Alfvénic units) is $\ell_0 \vect b$, where $\ell_0$ is a constant length. The equations below are written in the units of length where $\ell_0 = 1$; we come back to this normalization in Section~\ref{sec:twocorr}.
The time dependence comes from the evolution of the velocity field $\vect v(\vect r)$ and vector potential $\vect a(\vect r)$ following the MHD equation. We rearrange terms in this equation by combining  potential terms with the gradient of pressure 
 \begin{eqnarray}\label{NSEQ}
 && \vect \nabla \cdot \vect v = \vect \nabla \cdot \vect a =0;\\
    &&\partial_t \vect v = \vect v\times \vect \omega  -\nu \vect \nabla \times \vect \omega  \nonumber\\
    && \quad + (\vect b \cdot \vect \nabla) \vect b- \vect \nabla \left( p + \frac{\vect v^2 + \vect b^2}{2}\right);\\
    && \pd{t} \vect a = \vect v \times \vect b  - \eta \vect \nabla \times \vect b  - \vect \nabla\phi ;\\
    &&\vect \omega = \vect \nabla \times \vect v;\\
    && \vect b = \vect \nabla \times \vect a;
\end{eqnarray}
We restrict ourselves to three-dimensional Euclidean space, the most interesting case for physics applications. The generalization to arbitrary dimension is also possible, as discussed in previous papers \cite{M93, M23PR, migdal2023exact}.
The  circulations $\Gamma_v[C], \Gamma_a[C]$ satisfy the following evolution (with gradients of potentials dropping from the closed loop integrals)
\begin{subequations}\label{GammaEq}
    \begin{eqnarray}
    &&\pd{t} \Gamma_v[C] = \oint_{C} d \vect r \cdot \left( \vect v\times \vect \omega -\nu \vect\nabla \times \vect \omega\right.\nonumber\\
    &&\quad\left.  + (\vect b \cdot \vect \nabla) \vect b\right);\\
    &&\pd{t} \Gamma_a[C] =  \oint_{C} d \vect r \cdot \left(\vect v \times \vect b  - \eta \vect \nabla \times \vect b\right)
\end{eqnarray}
\end{subequations}

The loop functionals generate the correlation functions of vorticity by variations by the shape of the loop (so-called area derivative \cite{Mig83, M93, M23PR}). This relation between the loop functional and the correlation function was discussed in great detail in these review papers. Recently, we revisited the loop calculus \cite{migdal2025duality}, defining it with a polygonal approximation of the loop in the limit of the number of vertices $N \to \8$. This recent paper also discusses the Cauchy problem for the loop equations including the polygonal approximation.

In this paper, we use the language of continuum theory, where $\vect C(\theta)$ is a piecewise smooth loop in $\mathbb R_3$ and the momentum loop  $\vect P(\theta,t)$(see below) is a singular loop in $\mathbb R_3$ with the discontinuity at every angle $\theta$. Such bounded variation function ~\cite{Ambrosio2000,Volpert1967} can be defined either by slowly convergent Fourier series or as a limit of a polygon with $N \to \infty$ sides, with lengths $\Delta \vect P(\theta_k)$ staying finite in the limit $N \to \8$.

\section{High magnetic viscosity in strong magnetic field}

As a useful limiting case, consider an incompressible conducting fluid
subjected to a constant and uniform magnetic field \cite{okamoto2010}
\begin{equation}
    \boldsymbol{B}_0=B_0\boldsymbol{e}_z .
\end{equation}
We assume that the magnetic Reynolds number is small,
\begin{equation}
    \mathrm{Rm}=\frac{UL}{\eta}\ll 1,
\end{equation}
while the interaction parameter measuring the Lorentz force relative to
inertia need not be small. Writing the total magnetic field as
$\boldsymbol{B}=\boldsymbol{B}_0+\boldsymbol{b}$, the induced field satisfies
$|\boldsymbol{b}|/B_0=O(\mathrm{Rm})$. To leading order, the induction
equation therefore reduces to the quasi-static balance
\begin{equation}
    \eta\nabla^2\boldsymbol{b}
    +B_0\partial_z\boldsymbol{v}=0,
    \qquad
    \boldsymbol{b}
    =-\frac{B_0}{\eta}\nabla^{-2}\partial_z\boldsymbol{v}.
\end{equation}
Here $\nabla^{-2}$ is defined using the boundary conditions of the periodic
or unbounded domain, with the spatial zero mode removed. The leading
non-potential part of the Lorentz force is consequently
\begin{equation}
    (\boldsymbol{B}_0\cdot\nabla)\boldsymbol{b}
    =-\frac{B_0^2}{\eta}
      \frac{\partial_z^2}{\nabla^2}\boldsymbol{v},
\end{equation}
where the remaining gradient contribution has been absorbed into a
modified pressure.
In this case, the magnetic force is related to the instant value of the velocity field, which resulted in the following equation 
\begin{eqnarray}
    &&\pd{t}\vect v = \nonumber\\
    && \vect v \times \vect \omega - \nu \vect \nabla \times \vect \omega  - \vect \nabla \left(p + \frac{\vect v^2}{2}\right) 
    - \frac{B_0^2}{\eta \vect \nabla^2} \pd{z}^2 \vect v
\end{eqnarray}
This equation leads to the following loop equation (we use methods of \cite{migdal2024quantum,migdal2025duality} and operators $\hat v, \hat \nabla, \hat \omega$ defined in these papers):
\begin{eqnarray}
    &&-\I \nu\pd{t} \Psi_v(C,t) =  \oint d \theta \vect{C}'(\theta) \cdot\nonumber\\
    && \left(\hat v\times \hat \omega - \nu \hat \nabla \times \hat \omega -\frac{B_0^2}{\eta \hat \nabla^2} \hat \nabla_z^2 \hat v\right) \Psi_v(C,t);\\
    &&\hat \nabla = \ff{\vect C'(\theta-0)} -  \ff{\vect C'(\theta+0)};\\
    && \hat \omega = \I \nu \ff{\vect C'(\theta+0)} \times  \ff{\vect C'(\theta-0)};\\
    &&  \hat v = \frac{-1}{\hat \nabla^2} \hat \nabla \times \hat \omega
\end{eqnarray}
These operators are normalized so that they represent the gradient, the vorticity and the velocity at the point $\theta$ of the loop inside the average: $\hat \omega\, \Psi_v = \VEV{\vect \omega\, e^{\I \Gamma_v/\nu}}$, and likewise for $\hat v$ and $\hat \nabla$. Then the three terms in the loop equation are those of the equation for the velocity above.
The one-sided functional derivatives
$\ff{\vect C'(\theta\pm0)}$ should be understood as the two traces at a
jump of a bounded-variation (BV) momentum loop on the parameter circle
$S^{1}$. Their difference is therefore not a regularization artifact, but
the intrinsic jump derivative associated with velocity circulation as a
functional of the coordinate loop $\vect C(\theta)$. In the
path-ordered operator calculus (POOC), the two insertions are first kept
at distinct, cyclically ordered points $\theta_-<\theta_+$ on $S^{1}$;
the required antisymmetric products or commutators are formed at finite
separation, and only afterwards is the ordered coincidence limit
$\theta_\pm\to\theta$ taken. This ``order first, coincide second''
prescription preserves the information that would be lost in a naive
equal-point product and provides a geometric definition of the
operators entering the loop equation
\cite{MMEq79,Mig83,M23PR,migdal2025duality,migdal2025YangMills,
Migdal2026POOC}.

The loop Fourier transform maps the original functional dynamics to a
one-dimensional BV momentum loop $\vect P:S^{1}\to\mathbb{C}^{d}$.
At finite polygonal cutoff, the jump vectors are constrained to have
fixed length, so that the rescaled momentum loop takes values on a
compact spherical target. Closure of the loop then restricts the
allowed configurations to the compact manifold of closed equal-step
spherical polygons; its planar rational representatives form the Euler
ensemble.  The statistical
weight of these states is fixed by counting: every admissible state
enters once, and the count is dominated by the even-$N$ ensemble at
zero winding, which is also the only sector whose vorticity correlation
is nonnegative at zero separation for every state
\cite{MigdalQPTF2026}.

The physical loop calculus originated with the Makeenko--Migdal loop
equation \cite{MMEq79} and was subsequently developed and reviewed in
Refs.~\cite{Mig83,M23PR,migdal2025duality,migdal2025YangMills}.
Complementary mathematical work has placed the momentum-loop
construction and its BV limits on a more rigorous footing
\cite{DeLellisInprep}. In the following, we use the ordered BV
prescription described above, referring to
Ref.~\cite{MigdalQPTF2026} for its most recent formulation and for the
geometry and stability of the associated compact Euler ensemble.
In the momentum representation \cite{migdal2023exact,migdal2024quantum,MigdalQPTF2026}, the right side of this equation simplifies to a local rational function 
\begin{eqnarray}
   && \Psi_v(C,t) =\VEV{\exp{\I \oint d \theta \vect C'(\theta) \cdot \vect P(\theta,t)}}_P;\\
   \label{PeqB}
    &&\pd{t} \vect P(\theta,t) =- \nu \Delta \vect P \times (\vect P \times \Delta \vect P) \nonumber\\
    && \quad + \I\, \hat v\times (\vect P \times \Delta \vect P) 
    - \frac{B_0^2}{\nu \eta \Delta \vect P^2} (\Delta P_z)^2 \hat v;\\
    && \hat v= \frac{\nu \Delta \vect P \times (\vect P \times \Delta \vect P)}{\Delta \vect P^2} 
\end{eqnarray}
where $\Delta \vect P$ stands for the discontinuity $\Delta \vect P(\theta,t) = \vect P(\theta+0,t)- \vect P(\theta-0,t)$, and $\vect P$ on the right side for the half-sum of these two values. Without the magnetic term this is the momentum loop equation of the hydrodynamic turbulence \cite{migdal2023exact,MigdalQPTF2026}.

There are three possible regimes of asymptotic solutions of this equation
\subsection{anisotropic power decay}
In this solution, the momentum loop satisfies the constraint
\begin{eqnarray}
    \Delta P_z =0
\end{eqnarray}
which nullifies the last term (magnetic force). The solution is almost the same as in hydrodynamic turbulence:
\begin{eqnarray}
&& \vect P(\theta,t) = \frac{\vect F(\theta)}{\sqrt{ 2 \nu (t+ t_0)}};\\
   && 0 = (1- \Delta \vect F^2)\vect F + (\Delta \vect F\cdot \vect F)\Delta \vect F\nonumber\\
   && \quad + \I \left(\frac{(\Delta \vect F\cdot \vect F)^2}{\Delta \vect F^2} - \vect F^2\right)\Delta \vect F
\end{eqnarray}
This fixed point is the same as in the case of the hydrodynamic turbulence \cite{migdal2023exact, migdal2024quantum, migdal2025duality}, \textbf{except in this case, the global random rotation is replaced by the $O(2)$ rotation in $x-y$ plane}
\begin{eqnarray}\label{Fsol}
    && \vect{F}_k =  \frac{\left\{\cos (\alpha_k + \phi), \sin (\alpha_k + \phi), \I \cos(\beta/2)\right\}}{2 \sin \left(\frac{\beta }{2}\right)} ;\\
    && \theta_k = \frac{2 \pi k}{N}; \; \beta = \frac{2 \pi p}{q};\; N \to \infty;\\
    &&\alpha_{k} = \beta \sum_{l=1}^k \sigma_l;\; \sigma_l =\pm 1,\;\beta \sum_1^N\sigma_k = 2 \pi p r;\\
    && (\vect F_k - \vect F_{k+1})^2 = 1;\\
    && (\vect F_k + \vect F_{k+1})\cdot (\vect F_k - \vect F_{k+1})=0;\\
    && (\vect F_k + \vect F_{k+1})^2=0
\end{eqnarray}
The first two of these relations nullify the first two terms of the equation, and the third one nullifies its last term (the advection); this is what the imaginary third component of $\vect F_k$ is needed for.
The parameters $ p,q,r,\phi\in (0,2\pi),\sigma_0\dots\sigma_{N}= \sigma_0$ are random, making this solution for $\vect{F}(\theta)$ a fixed manifold rather than a fixed point.
We suggested in \cite{migdal2023exact} calling this manifold the big Euler ensemble or just the Euler ensemble. Later, we realized that this is a string theory with target space given by all regular star polygons, with extra Ising degrees of freedom on the sides of each polygon.

This solution corresponds to the vorticity correlation the same as in \cite{MigdalQPTF2026}, with integration over $\Omega \in  SO(3) $ replaced by integration over $O(2)$.
\subsection{exponential decay?}
Another potential solution would correspond to exponentially small $\vect P,\Delta  \vect P $ in which case only the magnetic term remains
\begin{eqnarray}
   && \pd{t} \vect P = -\frac{B_0^2}{\eta \Delta \vect P^2} (\Delta P_z)^2 
   \frac{\Delta \vect P \times (\vect P \times \Delta \vect P)}{\Delta \vect P^2} \nonumber\\
   && \quad =-\frac{B_0^2 (\Delta P_z)^2}{\eta \Delta \vect P^4} 
   \Delta \vect P \times (\vect P \times \Delta \vect P);
\end{eqnarray}
The exponential solution must satisfy linearized equation, with only the last term remaining in \eqref{PeqB}. This term yields a vector equation:
\begin{eqnarray}
    && \vect P(\theta, t) = \vect G(\theta) \exp{- \kappa t};\\
    && \kappa (\Delta \vect G)^4 \vect G = \frac{B_0^2}{\eta } (\Delta G_z)^2 \nonumber\\
    && \left(\Delta \vect G^2 \vect G - \Delta \vect G \left(\vect G\cdot \Delta \vect G\right)\right)
\end{eqnarray}
Comparing coefficients in front of $\vect G , \Delta \vect G$, we find two scalar equations
\begin{eqnarray}
   && \vect G \cdot \Delta \vect G =0;\\
   && \kappa   = \frac{B_0^2(\Delta G_z)^2}{\eta \Delta \vect G^2 } 
\end{eqnarray}
We see that the solution for $\kappa$ is positive: this is the Joule damping, with the decrement $\kappa = \frac{B_0^2}{\eta} \cos^2 \vartheta$, where $\vartheta$ is the angle between the discontinuity $\Delta \vect P$ and the magnetic field. This angle must be the same at all the discontinuities of the loop, as they decay with the same decrement.
The neglected terms are of the third degree in $\vect P$, so that they remain exponentially small compared to the magnetic term, and this exponential decay is consistent with our equation. It is not a turbulent regime, though: such a momentum loop tends to zero, its phase factor tends to one, and the motion it describes dies out in a time of the order of $1/\kappa$. The decrement vanishes at $\Delta P_z \to 0$; therefore, the motion that survives at large time brings us back to the previous case with the magnetic term vanishing and solution decaying as $ 1/\sqrt{t}$.
\subsection{fixed point?}
The third possibility is a fixed point: the time-independent solution of the momentum loop equation.  
In this case, the sum of all terms on the right side must vanish, which leads to an algebraic recurrent equation (with $k =1,\dots N$ labeling the vertices of the polygon $\vect P(\theta)$)
\begin{eqnarray}
     &&\nu \Delta \vect P \times (\vect P \times \Delta \vect P) = \nonumber\\
     && \quad \I\, \frac{\nu \Delta \vect P \times (\vect P \times \Delta \vect P)}{\Delta \vect P^2} \times (\vect P \times \Delta \vect P) \nonumber\\
     && \quad - \frac{(\vect B\cdot\Delta \vect P)^2}{\eta \Delta \vect P^2}  
     \frac{\Delta \vect P \times (\vect P \times \Delta \vect P)}{\Delta \vect P^2};\\
     && \Delta \vect P = \vect P_{k+1} - \vect P_{k};\\
     && \vect P = (\vect P_k + \vect P_{k+1})/2;
\end{eqnarray}
The first term on the right side (the advection) is directed along $\Delta \vect P$:
\begin{eqnarray}
   && \left(\Delta \vect P \times (\vect P \times \Delta \vect P)\right) \times (\vect P \times \Delta \vect P) = \nonumber\\
   && \quad -(\vect P \times \Delta \vect P)^2\, \Delta \vect P
\end{eqnarray}
The other two terms (the viscous and the Joule dissipation) are orthogonal to $\Delta \vect P$. Therefore, the two parts of the equation must vanish separately:
\begin{eqnarray}
   && (\vect P \times \Delta \vect P)^2 = 0;\\
   && \left(\nu + \frac{(\vect B\cdot\Delta \vect P)^2}{\eta\, (\Delta \vect P^2)^2}\right) \Delta \vect P \times (\vect P \times \Delta \vect P) = 0
\end{eqnarray}
The discontinuities $\Delta \vect P$ of the momentum loop must be real: the imaginary part of the loop is a constant vector, as in \eqref{Fsol}; otherwise, the phase of the loop functional would be complex for a real loop $C$. Then the factor in the last equation is positive, as the two kinds of dissipation add up, and we find $\vect P \times \Delta \vect P = 0$ at every discontinuity. This is the operator of vorticity: such a loop carries no vorticity.

We conclude that there are no fixed point solutions, which leaves us with decaying turbulence as the only alternative.
\subsection{Energy spectrum of  decaying turbulence in strong magnetic field}
The decaying turbulence regime reduces to an anisotropic Euler ensemble, with random rotation in the $xy$ plane instead of random rotation over whole $O(3)$.

In the isotropic case the vorticity correlation of the Euler ensemble was reduced in \cite{MigdalQPTF2026} to the Gibbs state of two quantum pendulums. It is given by the conditional source $C(h)=\widehat G(h)/\widehat Z(h)$, a ratio of two quadratures of the resolvents of the pendulums, taken at the separation projected on the plane of the polygon and then averaged over the orientations of this plane. Here the plane is fixed, normal to the magnetic field, so no average is left:
\begin{eqnarray}
\label{wwK}
    &&\VEV{\vect\omega(\vect 0) \cdot \vect \omega(\vect{r})} = \left(\frac{\nu}{\tilde\nu}\right)^{\oh}\frac{C\left(h\right)}{4 t^2};\\
    && h = \frac{|\vect r_\perp|}{\sqrt 2 L};\quad L = \sqrt{\tilde \nu t};\quad \tilde \nu = \nu N^2 \to \const{}
\end{eqnarray}
The source equals $C(0) = \sfrac{1}{8}$ at zero separation and decays as $1/h^2$ at large $h$, with a smaller oscillating term $\propto \cos (2 h + \mathrm{const})\, h^{-\sfrac{5}{2}}/\log h$.

The correlation does not depend on $z$, which leads to the expected delta function $\delta(k_{\parallel})$ in the longitudinal spectrum.
The collapse of a longitudinal part of the spectrum in an infinite system in a strong homogeneous magnetic field follows from translational invariance in the $z$ direction.

The remaining Fourier transform in the $xy$ plane is a Hankel transform, the same one that defines the spectral function $P(\kappa)$ of \cite{MigdalQPTF2026}:
\begin{eqnarray}
    &&\VEV{\vect\omega(\vect 0) \cdot \vect \omega(\vect{k})} \equiv \int d^3 \vect{r} \VEV{\vect\omega(\vect 0) \cdot \vect \omega(\vect{r})} \exp{- \imath \vect{k} \cdot \vect{r}} \nonumber\\
    &&\quad = \left(\frac{\nu}{\tilde\nu}\right)^{\oh}\frac{\pi^2 L^2}{t^2}\, \delta(k_{\parallel})\,\frac{P\left(|\vect k_\perp| L\right)}{|\vect k_\perp| L};\\
     && P(\kappa)=2 \kappa \int_0^\infty d h\, h\, J_0\left(\sqrt 2 h \kappa\right) C(h);
\end{eqnarray}
The transverse energy spectrum, normalized by $\oh \VEV{\vect v^2} = \int_0^\infty d k_\perp E(k_\perp,t)$, becomes
\begin{eqnarray}
    && E(k_\perp,t) = \left(\frac{\nu}{\tilde\nu}\right)^{\oh}\frac{\tilde{\nu}^{2}}{8 L}\,\frac{P\left(k_\perp L\right)}{(k_\perp L)^2}
\end{eqnarray}
\pctPDF{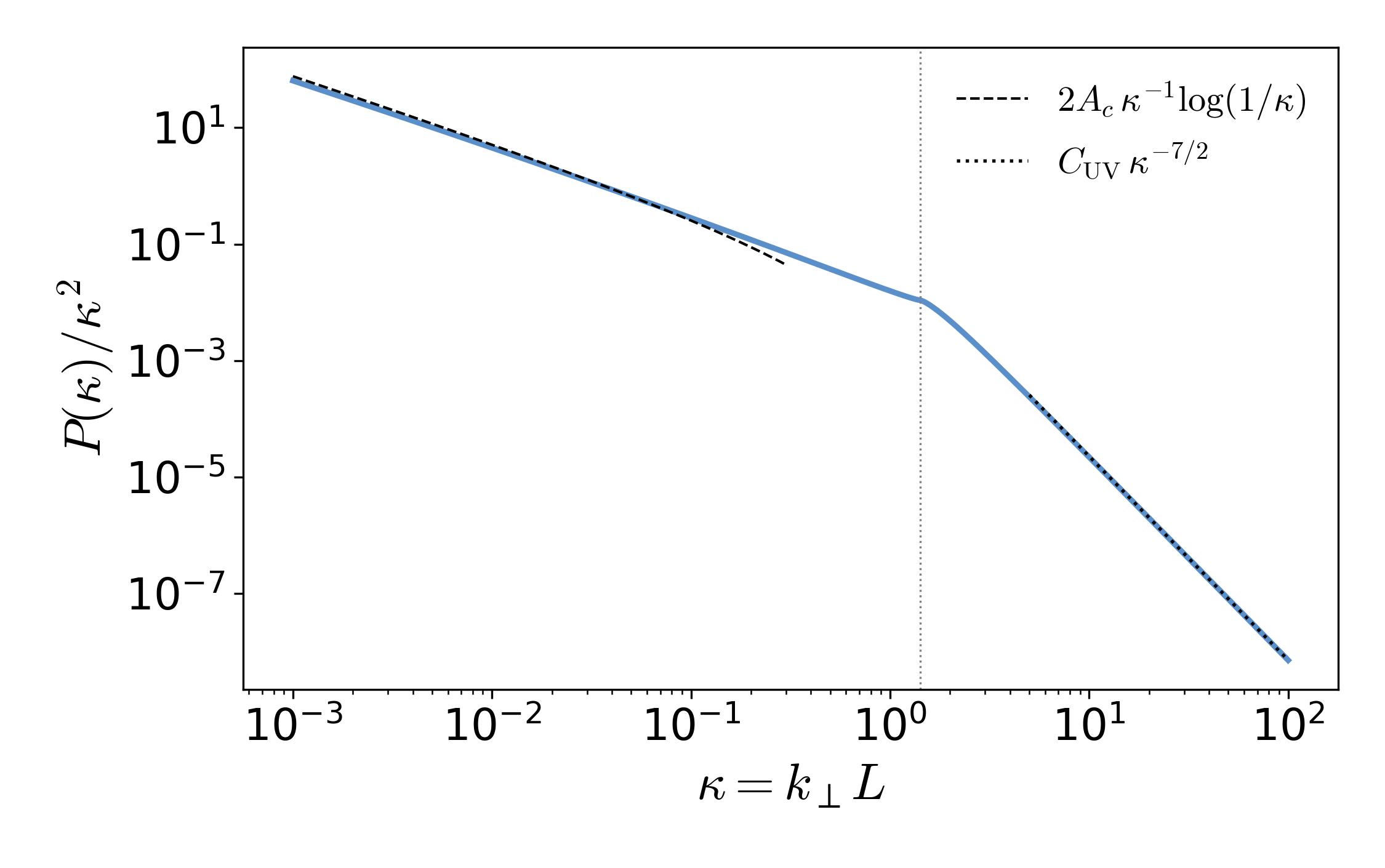}{The transverse energy spectrum $E(k_\perp,t)$ for MHD in strong magnetic field at large magnetic viscosity, in units of $(\nu/\tilde\nu)^{1/2}\tilde\nu^2/(8 L)$, as a function of scaling variable $\kappa =k_\perp L$ at fixed time $t$. Dashed line: the leading law $2 A_c \kappa^{-1}\log(1/\kappa)$ at small $\kappa$, with $A_c = 1/(128\sqrt 2)$; dotted line: the law $C_{\rm UV}\, \kappa^{-7/2}$ at large $\kappa$, with $C_{\rm UV} = 0.0722$; vertical dotted line: the threshold $\kappa = \sqrt 2$.}

As a function of the scaling variable $ |\vect k_\perp| \sqrt{\tilde \nu t}$ this is the spectrum of isotropic decaying turbulence \cite{MigdalQPTF2026} in the quenched order; the only difference is that the average over the orientation of the plane is absent, so that the two orders of this average discussed below coincide. The function $P(\kappa)$ is positive at all computed wavenumbers, with $\int_0^\infty P(\kappa) d \kappa = \sfrac{1}{8}$. The spectrum falls as $k_\perp^{-1} \log(1/k_\perp L)$ at $k_\perp L \ll 1$ and as $k_\perp^{-\sfrac{7}{2}}$ at $k_\perp L \gg 1$; between them, at $k_\perp L = \sqrt 2$, there is a threshold of the pendulums, where the spectrum is continuous and the slope of $P$ diverges as a double logarithm. The numerical table of the universal function $P(\kappa)$ is provided with \cite{MigdalQPTF2026}. The plot is shown in Fig.~\ref{fig::SpectrumPerp}.
\section{Loop Equation for full MHD turbulence}

Let us turn to the general case when both Reynolds numbers can go to infinity simultaneously at fixed ratio (Prandtl number).

 We are taking a more general Ansatz:
\begin{eqnarray}
  && \Psi[C_1, C_2, t] = \VEV{\exp{\I \frac{\Gamma_v(C_1)}{\nu} + \I \frac{\Gamma_a(C_2)}{\eta}}} = \nonumber\\
    && \quad\iint [D Q_1] [D Q_2]W[Q_1, Q_2,t]\exp{\I \mathcal A };\\
  && \mathcal A = \oint d \theta \left(\vect C_1'(\theta) \cdot \vect Q_1(\theta)  + \vect C_2'(\theta) \cdot \vect Q_2(\theta)\right);
\end{eqnarray}
The two loops are attached to a common base point, and each of them is restored from its tangent vector:
\begin{eqnarray}\label{basepoint}
    && \vect C_{1,2}(\theta) = \vect C_{1,2}(0) + \int_0^\theta d \theta' \vect C_{1,2}'(\theta');\\
    && \vect C_1(0) = \vect C_2(0)
\end{eqnarray}
Taking the time derivative, we find
\begin{eqnarray}
    &&-\I\pd{t}\Psi[C_1, C_2, t] = \nonumber\\
    && \quad\iint [D Q_1] [D Q_2]\left( -\I\pd{t} W[Q_1, Q_2,t] \right) \exp{\I\mathcal A};
\end{eqnarray}
According to the MHD equations \eqref{GammaEq} for the two circulations, this must be equal to
\begin{eqnarray}\label{PsiEq}
    &&-\I\pd{t}\Psi[C_1, C_2, t] =\nonumber\\
    && \quad\iint [D Q_1] [D Q_2]W[Q_1, Q_2,t]\oint d \theta \exp{\I \mathcal A}\nonumber\\
    &&\left(\nu\, \vect C_1'(\theta) \cdot \left(\hat v_1\times\hat \omega_1 - \hat\nabla_1 \times \hat \omega_1\right) \right.+ \nonumber\\
    && \quad\frac{\eta^2}{\nu}\, \vect C_1'(\theta) \cdot (\hat \nabla_1 \times \hat b_1) \times \hat b_1+ \nonumber\\
    && \quad\left.\vect C_2'(\theta) \cdot \left( \nu\,\hat v_2 \times \hat \omega_2-\eta \hat \nabla_2 \times \hat \omega_2\right)\right);
\end{eqnarray}
Here the operators with hats represent the fields at the point $\theta$ of the loop in the units of the two viscosities: $\hat v = \vect v/\nu$, $\hat \omega = \vect \omega/\nu$ and $\hat b = \vect b/\eta$. These units lead to the coefficients in this equation. The vorticity and the magnetic field are generated by the area derivatives of their loops:
\begin{eqnarray}\label{omegaops}
    && \hat \omega_1 = \I\,\vect Q_1\times\Delta\vect Q_1;\\
    && \hat b \equiv \hat\omega_2 = \I\,\vect Q_2\times\Delta\vect Q_2
\end{eqnarray}
The magnetic-field operator is in exact analogy with the vorticity operator. The magnetic force is written here as $(\vect \nabla \times \vect b) \times \vect b$. It differs from $(\vect b \cdot \vect \nabla) \vect b$ in \eqref{GammaEq} by the gradient of $\vect b^2/2$, which drops from the circulation; in this form the force involves the field and its curl, like the term $\vect v \times \vect \omega$.

The remaining problems are:
\begin{itemize}
    \item translate the vector $\vect b(\vect r)$  from $ \vect r = \vect C_2(\theta)$ where it corresponds to the operator $\hat b(\theta) = \I \vect Q_2 (\theta)\times \Delta \vect Q_2(\theta)$ to a point $\vect r = \vect C_1(\theta)$.
    \item translate the vector $\vect v(\vect r)$  from $ \vect r = \vect C_1(\theta)$ where it corresponds to the operator $\hat v_1(\theta)$ to a point  $\vect r = \vect C_2(\theta)$.
\end{itemize}
\subsection{Translation operator  and magnetic force}
The translation operators $\hat T_{1,2}$ can be related to the gradient operator, which yields the discontinuity $\Delta \vect Q$ when applied to the exponential in the MLE
\begin{eqnarray}
&&\vect \nabla_{1,2}(\theta) \Rightarrow \int_{\theta-0}^{\theta+0} d \theta'\ff{\vect C_{1,2}(\theta')} \Rightarrow\nonumber\\
    && \quad -\I\int_{\theta-0}^{\theta+0} d \theta' \vect Q'_{1,2}(\theta') =-\I\Delta \vect Q_{1,2}(\theta);
\end{eqnarray}
Accordingly, in this section and in the next one $\hat \nabla_{1,2} = -\I\Delta \vect Q_{1,2}$, and the velocity operator is given by the \BS{} law, $\hat v_1 = -\hat \nabla_1 \times \hat \omega_1/\hat \nabla_1^2$.
Using this representation, we find:
\begin{eqnarray}
    &&\hat T_1 = \exp{-\I \left(\vect C_1(\theta) - \vect C_2(\theta)\right)\cdot \Delta \vect Q_2(\theta)};\\
    && \hat T_2 = \exp{-\I \left(\vect C_2(\theta) - \vect C_1(\theta)\right)\cdot \Delta \vect Q_1(\theta)};
\end{eqnarray}
This shift only relates to the magnetic force and the $\hat v_2\times\hat \omega_2$ terms in the equation; let us single these terms out, starting with the magnetic force term:
\begin{eqnarray}
&& \VEV{ \left[(\vect \nabla \times \vect b) \times \vect b\right]_{\vect C_1(\theta)}\exp{\I \frac{\Gamma_v}{\nu} + \I \frac{\Gamma_a}{\eta}}} =\nonumber\\
&& \VEV{\exp{\left(\vect C_1(\theta) - \vect C_2(\theta)\right)\cdot \pd{\vect C_2(\theta)} } \right.\nonumber\\
    && \quad\left.\left[(\vect \nabla \times \vect b) \times \vect b\right]_{\vect C_2(\theta)}\exp{\I \frac{\Gamma_v}{\nu} + \I \frac{\Gamma_a}{\eta}}} =\nonumber\\
    && \eta^2\iint [D Q_1] [D Q_2]W[Q_1, Q_2,t]\nonumber\\
    && \quad\hat T_1\left(\hat \nabla_2 \times \hat \omega_2\right) \times \hat \omega_2 \exp{\I \mathcal A}=\nonumber\\
    && \eta^2\iint [D Q_1] [D Q_2]W[Q_1, Q_2,t] \nonumber\\
    && \quad \left(\hat \nabla_2 \times \hat \omega_2\right) \times \hat \omega_2 \exp{\I \mathcal B_1(\theta)};\\
    &&\mathcal B_1(\theta) =\mathcal A + \left(\vect C_2(\theta) - \vect C_1(\theta)\right)\cdot \Delta \vect Q_2(\theta)
\end{eqnarray}
By \eqref{basepoint}, the extra term in the exponential corresponds to shifts
\begin{eqnarray}\label{shiftsOne}
    &&\vect Q_1(\theta') \Rightarrow \vect Q_1(\theta') - \Delta\vect Q_2(\theta) \Theta(\theta - \theta');\\
    &&\vect Q_2(\theta') \Rightarrow \vect Q_2(\theta') + \Delta\vect Q_2(\theta) \Theta(\theta - \theta');
\end{eqnarray}
on the arc of the parametric circle between the base point and the point $\theta$. After these shifts the magnetic loop has no discontinuity at the point $\theta$, and the discontinuity of the hydrodynamic loop at this point becomes $\Delta \vect Q_1(\theta) + \Delta \vect Q_2(\theta)$. This is how the translation acts in the space of momentum loops: it hands over the jump of the momentum from one loop to another. The shifts also create the opposite jumps at the base point, $-\Delta \vect Q_2(\theta)$ in the hydrodynamic loop and $+\Delta \vect Q_2(\theta)$ in the magnetic one, so that the sum of the jumps of each loop remains zero; these two jumps cancel in the phase, because the two loops are attached to the same point \eqref{basepoint}. The phase $\mathcal B_1(\theta)$ is the phase $\mathcal A$ of another pair of momentum loops, different for every $\theta$.

The shifts \eqref{shiftsOne} are proportional to the integration variable $\Delta\vect Q_2(\theta)$ itself. They make a linear transformation of the pair $\vect Q_1, \vect Q_2$, and this transformation is a projection: it annihilates the jump $\Delta\vect Q_2(\theta)$. Therefore, it cannot be undone by a shift of the integration variables in the functional integral, and the translated term cannot be reduced to the terms with the phase $\mathcal A$.
\subsection{Translation operator and advection of vector potential}
Likewise, the advection term in the magnetic circulation
\begin{eqnarray}
    &&\iint [D Q_1] [D Q_2]W[Q_1, Q_2,t]\nonumber\\
    &&\quad\oint d \theta\,\vect C_2'(\theta) \cdot  \hat T_2\left(\hat v_1 \times \hat \omega_2 \right) \exp{\I \mathcal A}=\nonumber\\
    && \iint [D Q_1] [D Q_2]W[Q_1, Q_2,t]\nonumber\\
    &&\quad\oint d \theta\,\vect C_2'(\theta) \cdot \left(\hat v_1 \times \hat \omega_2\right) \exp{\I \mathcal B_2(\theta)};\\
    && \mathcal B_2(\theta) = \mathcal A + \left(\vect C_1(\theta) - \vect C_2(\theta)\right)\cdot \Delta \vect Q_1(\theta)
\end{eqnarray}
The extra term in the exponential corresponds to shifts
\begin{eqnarray}\label{shiftsTwo}
    &&\vect Q_1(\theta') \Rightarrow \vect Q_1(\theta') + \Delta\vect Q_1(\theta) \Theta(\theta - \theta');\\
    &&\vect Q_2(\theta') \Rightarrow \vect Q_2(\theta') - \Delta\vect Q_1(\theta) \Theta(\theta - \theta');
\end{eqnarray}
Now the hydrodynamic loop has no discontinuity at the point $\theta$, and the discontinuity of the magnetic loop becomes $\Delta \vect Q_1(\theta) + \Delta \vect Q_2(\theta)$. This is the same projection, with the two loops exchanged.
\subsection{Loop equation in momentum loop space}
Collecting the terms, we find
\begin{eqnarray}\label{fullMagEq}
    &&-\I\pd{t}\Psi[C_1, C_2, t] =\nonumber\\
    && \quad\iint [D Q_1] [D Q_2]W[Q_1, Q_2,t]\oint d \theta\nonumber\\
    &&\Big(\nu\, \vect C_1'(\theta) \cdot \left(\hat v_1\times\hat \omega_1 - \hat\nabla_1 \times \hat \omega_1\right)\exp{\I \mathcal A} - \nonumber\\
    && \quad\eta\, \vect C_2'(\theta) \cdot \hat \nabla_2 \times \hat \omega_2\,\exp{\I \mathcal A} +\nonumber\\
    && \quad\nu\, \vect C_2'(\theta) \cdot \hat v_1 \times \hat \omega_2\,\exp{\I \mathcal B_2(\theta)} +\nonumber\\
    && \quad\frac{\eta^2}{\nu}\, \vect C_1'(\theta) \cdot \left(\hat \nabla_2 \times \hat \omega_2\right) \times \hat \omega_2\,\exp{\I \mathcal B_1(\theta)}\Big);
\end{eqnarray}
In the first two terms, the integration loop variables $\vect C_{1,2}'(\theta) $ are equivalent to functional derivatives $ \I \ff{\vect Q_{1,2}(\theta)}$ acting on the rest of factors in the integral, not counting the exponential (the functional integral of the functional derivative equals zero, so we could apply this functional derivative with opposite sign to remaining factors).
These two terms alone would lead to the \LI{} equation for the weight function $W$:
\begin{eqnarray}\label{LiouvilleEq}
    && -\pd{t} W[Q_1, Q_2,t] =\nonumber\\
     && \nu\oint d \theta\ff{\vect Q_1(\theta)} \cdot \left( \hat v_1 \times \hat \omega_1 - \hat\nabla_1 \times \hat \omega_1 \right)W[Q_1, Q_2,t] -\nonumber\\
     &&\eta\oint d \theta\ff{\vect Q_2(\theta)}\cdot \hat \nabla_2 \times \hat \omega_2\, W[Q_1, Q_2,t];
\end{eqnarray}
The last two terms of \eqref{fullMagEq} add to this equation the weights of the shifted loops \eqref{shiftsTwo}, \eqref{shiftsOne}, which have no discontinuity at the point $\theta$ in one of the two loops.
\section{Dimension reduction}
This functional integro-differential equation can be interpreted as \Schr{} equation in double loop space, which is a simplification compared to the Hopf equation. Still, one can go one giant step further in simplifying the MHD problem. Like in the case of hydrodynamics \cite{migdal2025duality}, one can reduce this equation to the set of singular one-dimensional equations.
\subsection{Collapse of the distribution}
We look for the solution collapsing on the two trajectories $Q_{1,2}(\theta, t)$, which means the functional delta function for the general solution $W$
\begin{eqnarray}\label{collapse}
   && W[Q_1, Q_2,t] \to \nonumber\\
   &&\delta[Q_1(.) - Q_1(.,t)]\delta[Q_2(.) - Q_2(.,t)]
\end{eqnarray}
The \LI{} equation \eqref{LiouvilleEq} is of the first order in functional derivatives. It is solved by such a delta function, moving with the velocity given by the coefficients of this equation:
\begin{eqnarray}\label{Heq}
  &&\pd{t} \vect Q_1(\theta,t)  = \nu \left(\hat v_1 \times \hat \omega_1 - \hat\nabla_1 \times \hat \omega_1\right);\\
\label{Meq}
  &&\pd{t} \vect Q_2(\theta,t)  = -\eta\, \hat\nabla_2 \times \hat \omega_2;
\end{eqnarray}
One can also see this by representing the functional delta function as a limit of the Gaussian with the variance $\lambda \to 0+$ and balancing the leading singular terms, as in \cite{migdal2025duality}. The first equation is the momentum loop equation of the hydrodynamic turbulence. The second one involves only the magnetic loop: this is the diffusion of the vector potential.

Let us now turn to the two translated terms in \eqref{fullMagEq}. With the collapsed weight \eqref{collapse}, each of them is equal to its coefficient, taken on the trajectory, times the phase factor $\exp{\I \mathcal B_{1,2}(\theta)}$. As we have seen, these are the phase factors of other momentum loops, different for every $\theta$. The phase factors of different momentum loops are linearly independent as functionals of the two loops $C_1, C_2$. Therefore, these terms cannot be balanced by the time derivative of the weight, and they cannot cancel each other. Each of the two coefficients must vanish:
\begin{subequations}\label{constraints}
\begin{eqnarray}
    && \hat v_1 \times \hat \omega_2 =0;\\
    && \left(\hat \nabla_2 \times \hat \omega_2\right) \times \hat \omega_2 = -\hat \nabla_2\, \hat \omega_2^2 = 0
\end{eqnarray}
\end{subequations}
which must be valid identically for every angle $\theta$. (In the Gaussian representation, the weights of the shifted loops are centered at other trajectories, at a finite distance from the solution, where the weight $W$ and its time derivative are exponentially small in the limit $\lambda \to 0$; therefore, nothing can balance these terms there.)

These are the constraints on the solution. The magnetic field operator must be parallel to the velocity operator at the same point of the loop, and it must be a null vector, which makes the magnetic field force-free. Under these constraints the magnetic variables do not enter the equation \eqref{Heq} for the hydrodynamic loop.

The same applies to a statistical ensemble of such trajectories, as long as the shifted loops do not belong to this ensemble. This is the case for the polygons considered below, where the discontinuities $\Delta \vect Q_{1,2}$ at all the vertices are finite.
\subsection{Gibbs distribution as initial data}
As initial data we take the Gibbs distribution\footnote{If mathematicians suggest ignoring thermal fluctuations and employing smooth initial data (the idealized Cauchy problem), I would remind them that the temperature in the Universe is never exactly zero. In hydrodynamic turbulence, the temperature is at least the freezing point of water ($0^\circ\mathrm{C}$), while in astrophysical turbulence, the cosmic microwave background ensures a minimal temperature of about $2\,\mathrm{K}$. Hence, the Gibbs distribution provides a physically correct description of initial data for the turbulent flows.}
\begin{eqnarray}\label{GibbsMHD}
    &&P_0[\vect v, \vect a]  = \exp{-\int d^3 r  \frac{\vect v^2 + \vect b^2}{2 T_0}};\\
    && \vect b = \nabla \times \vect a;
\end{eqnarray}
The corresponding initial value of the loop functional factorizes as follows:
\begin{eqnarray}
    &&\Psi[C_1, C_2;0] = \Psi_1[C_1] \Psi_2[C_2];\\
    && \Psi_1[C] \propto \exp{-m_0 |C|}; \, m_0 = \frac{T_0}{2 r_1^2 \nu^2};\\
    && \Psi_2[C] \approx \exp{-m_1 |C|\log(|C|/r_2)};\\
    && m_1 = \frac{T_0}{4 \pi \eta^2};
\end{eqnarray}
Here $|C|$ is the length of $C$. $r_1, r_2$ are some molecular length parameters, describing the variance of the thermal fluctuation distribution. 

In the Gibbs distribution, the correlations of fluctuations of velocity and magnetic field are assumed local. In real world, with the molecular structure of plasma, there will be some radii of correlations of these fluctuations.

\subsection{Initial value of hydro loop and relativistic particle} 
The computations are performed in Appendix.
The distribution for the hydro momentum loop $\vect Q_1$ was already computed in \cite{migdal2025duality}, we just repeat it using different notations, involving the temperature. The resulting distribution in the continuum limit $r_1 \to 0$ is Gaussian, plus the random proper time.

\begin{eqnarray}\label{Hloop}
    &&W[Q_1] \propto \int_0^\infty d T \nonumber\\
    &&\quad\exp{- \int_0^T d s\left(m^2 +\vect Q_1(s)^2\right)};\\
    && m^2 \propto \mu m_0^2
\end{eqnarray}
Here $\mu  \to 0$ is the decrement of the distribution $ \exp{-\mu N}$ of the number $N$ of vertices in the polygonal approximation of the loop $C$.
At finite temperature $T_0$ and $\mu \to 0, r_1 \to 0 $ this physical mass $m$ stays finite.

\subsection{Initial value of magnetic loop and QED in 3D} 
The computation of the initial distribution of the magnetic momentum loop is less trivial. 

Here, the independent variables are components of the vector potential $\vect a$. The gauge condition is irrelevant as the magnetic loop is gauge--invariant.
The Gibbs distribution for the magnetic energy reads (in our units for the vector potential)
\begin{eqnarray}
   W_0[a] \propto \exp{-\frac{\int d^3 \vect r (\vect \nabla \times \vect a)^2}{2 T_0}}
\end{eqnarray}

In the next section, we advance further, by finding an asymptotic solution of the MHD turbulence, given by the fixed trajectory of this ODE.
\section{Decaying MHD turbulence}
In absence of external forces, pumping in the energy to MHD, the turbulence kinetic energy is expected to dissipate in the small-scale vortex structures, similar to the decaying hydrodynamic turbulence.

The time decay of the full MHD turbulence in our theory is described by a scaling solution:
\begin{eqnarray}
   &&\vect Q_1(\theta,t)\to (2\nu(t+ t_0))^{-\oh} \vect f(\theta) ;\\
    && \vect Q_2(\theta,t)\to (2\nu(t+ t_0))^{-\oh} \vect g(\theta) ;
\end{eqnarray}
We shall use the polygonal regularization
\begin{eqnarray}
    &&\vect f_k = \vect f(2 \pi  k/N);\\
    &&\vect g_k = \vect g(2 \pi  k/N);
\end{eqnarray}
The right sides of the equations \eqref{Heq}, \eqref{Meq} are homogeneous functions of the third degree, and the constraints \eqref{constraints} are homogeneous as well. Therefore, all the terms of the equations scale as $(\nu(t+ t_0))^{-\sfrac{3}{2}} $, and we find two sets of coupled algebraic equations, generalizing the hydrodynamic MLE:
\begin{subequations}\label{MLE}
\begin{eqnarray}
 && \hat v_1\times\hat \omega_2 =0;\quad \hat \omega_2^2 =0;\\
    && \vect f_{k+} = \hat v_1 \times \hat \omega_1 -\hat\nabla_1 \times \hat \omega_1;\\
    &&  \vect g_{k+} = -\frac{\eta}{\nu} \hat \nabla_2 \times \hat \omega_2;
\end{eqnarray}
\end{subequations}

We use the following notations here:
\begin{subequations}\label{MLEdef}
\begin{eqnarray}
    && \vect f_{k+} = \frac{\vect f_k + \vect f_{k+1}}{2};\\
    && \Delta \vect f_k = \vect f_{k+1} - \vect f_{k};\\
    && \hat \omega_1 = \I \vect f_{k+} \times \Delta \vect f_k;\\
    && \hat \nabla_1 = \I \Delta \vect f_k;\\
    && \hat v_1 = \frac{-\hat \nabla_1 \times \hat \omega_1} {\hat \nabla_1^2};\\
    && \vect g_{k+} = \frac{\vect g_k + \vect g_{k+1}}{2};\\
    && \Delta \vect g_k = \vect g_{k+1} - \vect g_{k};\\
    && \hat \omega_2 = \I \vect g_{k+} \times \Delta \vect g_k;\\
    && \hat \nabla_2 = \I \Delta \vect g_k;
\end{eqnarray}
\end{subequations}
Here the operator $\hat \nabla$ is defined as $\hat \nabla = \I \Delta$, with the sign opposite to that of the previous sections. This changes the sign of $\hat v_1$ and of the right sides of the last two equations \eqref{MLE}, and it does not affect the constraints.

The first line of \eqref{MLE} is the constraints \eqref{constraints}. The equation for $\vect f_{k+}$ does not involve the magnetic polygon: this is the MLE of the hydrodynamic turbulence \cite{migdal2023exact, migdal2024quantum}. The equation for $\vect g_{k+}$ has no advection term.

These are recurrent algebraic relations for the $\vect f_{k}, \vect g_k$ given previous values $\vect f_{k-1}, \vect g_{k-1}$.  However, only periodic solutions are physically acceptable.

The periodicity requirement is a strong restriction of the space of possible solutions. For the hydrodynamic polygon $\vect f_k$ we take the known periodic solution, the Euler ensemble \cite{migdal2023exact, migdal2024quantum}: a random walk on a regular star polygon. With this $\vect f_k$, the remaining equations \eqref{MLE} for the magnetic polygon have an exact solution (up to the global $O(3)$ rotation $ \vect f_k \Rightarrow \hat \Omega \cdot \vect f_k,\vect g_k \Rightarrow \hat \Omega \cdot \vect g_k $):
\begin{subequations}\label{fgsolution}
\begin{eqnarray}
    &&f_k=\frac{\left\{\cos (\alpha _k),\sin (\alpha _k),\I \cos\left(\frac{\beta }{2}\right)\right\}}{2 \sin \left(\frac{\beta }{2}\right)};\\
    \label{aligned}
    &&g_k=\pm\sqrt{\Pr}\, f_k;\\
    &&\alpha_k = \beta\sum_{l=1}^k \sigma_l;\\
    && \sigma_l = \pm 1;\\
    && \beta = \frac{2 \pi p}{q};\\
    && \Pr = \frac{\nu}{\eta}
\end{eqnarray}
\end{subequations}
This solution is geometrically equivalent to synchronized random walks on a pair of regular star polygons: the magnetic polygon is the hydrodynamic one, rescaled by the factor $\sqrt{\Pr}$. It exists for every Prandtl number. The relation \eqref{aligned} holds for all three components of the vectors: the magnetic polygon has the same constant imaginary normal component as the hydrodynamic one, multiplied by the same factor, $\pm\sqrt{\Pr}\,\frac{\I}{2} \cot\left(\frac{\beta}{2}\right)$. In other words, $\vect Q_2 = \pm \vect f/\sqrt{2\eta(t+t_0)}$: the magnetic momentum loop is the hydrodynamic one, with the magnetic viscosity $\eta$ in place of $\nu$.

The verification is elementary. The vertices of the polygon $\vect f_k$ lie on a complex sphere, $\vect f_k^2 = \sfrac{1}{4}$, and its sides have a unit length, $(\Delta \vect f_k)^2 = 1$. Therefore,
\begin{eqnarray}\label{nullmid}
    \vect f_{k+} \cdot \Delta \vect f_k = 0;\quad \vect f_{k+}^2 = 0
\end{eqnarray}
The midpoint of every side is a null vector, orthogonal to this side. For such a vector $\Delta \vect f_k \times (\vect f_{k+} \times \Delta \vect f_k) = \vect f_{k+}$, and we find
\begin{eqnarray}\label{vertexops}
    && \hat\nabla_1 \times \hat \omega_1 = -\vect f_{k+};\quad \hat v_1 = -\vect f_{k+};\\
    && \hat \omega_1 = \sigma_{k+1} \vect f_{k+}
\end{eqnarray}
The advection term $\hat v_1 \times \hat \omega_1$ vanishes, as in the hydrodynamic Euler ensemble, and the equation for $\vect f_{k+}$ is satisfied. The operators of the polygon \eqref{aligned} are $\hat \omega_2 = \Pr\, \hat \omega_1$ and $\hat \nabla_2 = \pm \sqrt{\Pr}\, \hat \nabla_1$. The equation for $\vect g_{k+}$ reduces to the equation for $\vect f_{k+}$ without the advection term, and both constraints hold, because $\hat \omega_2$ is a null vector parallel to $\hat v_1$.

The normal component of the magnetic polygon is necessary here, for the same reason as the normal component of $\vect f_k$ is necessary in the hydrodynamic equation. For a polygon with $\vect f_k^2 = \vect f_{k+1}^2$, and likewise for $\vect g_k$, the definitions \eqref{MLEdef} yield
\begin{eqnarray}\label{nullmidpoints}
    && \hat v_1 \times \hat \omega_1 = \vect f_{k+}^2\, \hat \nabla_1;\\
    && \left(\hat \nabla_2 \times \hat \omega_2\right) \times \hat \omega_2 = (\Delta \vect g_k)^2\, \vect g_{k+}^2\, \hat \nabla_2
\end{eqnarray}
The magnetic force has the structure of the advection term, with the operator of the vector potential, $-\hat \nabla_2 \times \hat \omega_2/\hat \nabla_2^2$, in place of the velocity. Each of the two terms vanishes only if the midpoint of the side is a null vector. For a regular polygon in a plane, with the radius $\rho$ and the center at the origin, shifted along the normal by a constant $c$, the square of this midpoint is $\rho^2 \cos^2(\beta/2) + c^2$. Therefore, $c = \pm \I \rho \cos(\beta/2)$, which is $\pm\frac{\I}{2}\cot(\beta/2)$ for $\vect f_k$ and $\pm\frac{\I}{2}\sqrt{\Pr}\cot(\beta/2)$ for $\vect g_k$. The equation for $\vect g_{k+}$ alone is satisfied when $(\Delta \vect g_k)^2 = \Pr$ and $\vect g_k^2 = \vect g_{k+1}^2$, so that it holds with any constant normal component. The second constraint fixes this component up to the sign, and the first one admits only the turns by $0$ and $\pi$ of the magnetic polygon in its plane, each with its sign of the normal component, which is \eqref{aligned}. Like the normal component of $\vect f_k$, this is a constant vector, and it does not contribute to the phase of a closed loop in \eqref{solution}.

Thus, on this solution the operators of the velocity, the vorticity, the magnetic field and its curl at every vertex of the polygon are parallel to one null vector $\vect f_{k+}$. All the nonlinear terms of the MHD equations \eqref{NSEQ} vanish: $\vect v \times \vect \omega$, $\vect v \times \vect b$ and the magnetic force. The velocity is a \BL{} field, with the sign $\sigma_{k+1}$ of the step of the random walk, and the magnetic field is force-free and aligned with the velocity. Such a depression of nonlinearity is known in the direct numerical simulations of decaying MHD turbulence, where local \BL{} velocity fields, force-free magnetic fields and directional alignment of the velocity and the magnetic field emerge rapidly, in spatial patches \cite{Servidio2008, Matthaeus2008}.

With the given hydrodynamic polygon, \eqref{aligned} is the only magnetic polygon with a nonvanishing field and a real phase in the loop functional. Indeed, the equation for $\vect g_{k+}$, multiplied by $\Delta \vect g_k$, yields $\vect g_{k+} \cdot \Delta \vect g_k = 0$, after which it requires $(\Delta \vect g_k)^2 = \Pr$ or $\vect g_{k+} = 0$. By the first constraint, the vector $\vect g_k \times \vect g_{k+1}$ is parallel to the null vector $\vect f_{k+}$. If it does not vanish, both $\vect g_k$ and $\vect g_{k+1}$ are orthogonal to $\vect f_{k+}$, so that they lie in the plane of $\vect f_k$ and $\vect f_{k+1}$. The planes of two adjacent sides intersect along $\vect f_k$; therefore, $\vect g_k = \lambda_k \vect f_k$, and $(\Delta \vect g_k)^2 = \Pr$ leaves $\lambda_k = \lambda_{k+1} = \pm\sqrt{\Pr}$. There are two exceptions: at a turning point of the random walk, $\vect f_{k+1} = \vect f_{k-1}$, the two planes coincide, and $\vect g_k$ can be shifted along the null vector $\vect f_{k+}$; and the field vanishes on a side with $\vect g_{k+1} = - \vect g_k$. In the polygons using these exceptions the imaginary part of $\vect g_k$ varies along the polygon, unless the field vanishes on all sides or the whole polygon is shifted by a constant vector (the latter happens only for the two alternating sequences $\sigma_k = \pm(-1)^k$, and it does not change the loop functional). Then the phase in \eqref{solution} is complex for a real loop $C_2$, and we discard such solutions.

Classifying all periodic solutions of the hydrodynamic loop equation is a challenging mathematical task beyond our current goals. In theoretical physics, we instead focus on finding solutions with physically acceptable properties and validating them through experiments and numerical simulations. The significance of the particular periodic solution presented here should be judged by its agreement with physical data and numerical simulations rather than general existence or uniqueness theorems.

In the turbulent limit the two constraints \eqref{constraints} are enforced with different strength. In this limit the relevant polygons are the zig-zags of the critical layer $|\beta - \pi| = \xi/\sqrt{N}$ \cite{MigdalQPTF2026}, so that $\cot^2\left(\frac{\beta}{2}\right) \to \frac{\xi^2}{4 N}$, and the operators of all the fields at a vertex are small. Let us turn the magnetic polygon \eqref{aligned} in its plane by an arbitrary angle $\phi$, keeping its normal component. The midpoints of its sides remain null vectors. Therefore, the magnetic force vanishes, the equation for $\vect g_{k+}$ holds, and only the first constraint is violated:
\begin{eqnarray}\label{softmode}
    \left|\hat v_1\times\hat \omega_2\right| = \frac{\Pr}{4} \cot^2\left(\frac{\beta}{2}\right) \sqrt{\sin^2\phi + 4 \sin^2\frac{\phi}{2}}
\end{eqnarray}
This term enters the loop equation \eqref{fullMagEq} with the phase factor $\exp{\I \mathcal B_2(\theta)}$ of the shifted loops. Relative to the time derivative of the loop functional, its contribution decreases as $N^{-\oh}$ for two loops at a finite distance from each other, as well as for two coinciding loops, and it stays finite only for two loops shifted against each other by a distance of the order of $\sqrt{2 \nu t}$; we verified this numerically for $N$ up to $16384$. Thus, the term $\vect v \times \vect b$ of a turned polygon vanishes only in the turbulent limit, and not for the loops shifted against each other by the viscous scale. In this sense the loop equation is satisfied asymptotically at every turn $\phi$ and at every Prandtl number. The functional of each of the two loops, and with it the spectra and the energies of the two fields, does not depend on $\phi$. The alignment of the magnetic field with the velocity is fixed at the viscous scale, and the turn $\phi$ is a soft mode of the turbulent limit. Its fluctuations belong to the stability problem discussed below.

The second constraint is not soft. The term of the magnetic force enters \eqref{fullMagEq} with the coefficient $\eta^2/\nu$ and with two more powers of the momentum than the term $\hat v_1 \times \hat \omega_2$: it is larger by a factor of the order of $N^2$ times the ratio of the magnetic energy to the kinetic one. On the polygon \eqref{aligned} without its normal component, $\hat \omega_2^2 = -\frac{\Pr^2}{4} \cot^2\left(\frac{\beta}{2}\right)$, and the contribution of this term does not decrease in the turbulent limit, unless the magnetic energy is negligible. The magnetic field of the solution is force-free at every scale.

In summary, our asymptotic solution for decaying MHD turbulence takes the form
\begin{eqnarray}\label{solution}
    &&\Psi[C_1, C_2, t] = \lim_{N \to \8}\\
&&\VEV{\exp{\displaystyle\frac{\I\sum_{k=1}^N \Delta \vect C_1(\theta_k) \cdot \vect f_k +\Delta \vect C_2(\theta_k)\cdot \vect g_k }{\sqrt{2\nu t}} }}_{\mathcal E};\nonumber\\
&& \theta_k = 2 \pi k/N;
\end{eqnarray}
The magnetic force vanishes on this solution, so that it produces no corrections to it. The stability of this solution with respect to the perturbations violating the constraints \eqref{constraints} is a separate problem, which we leave for future work.

A formal continuum representation of this solution follows from the analysis of the Euler ensemble in \cite{migdal2024quantum}
\begin{subequations}\label{pathIntegralSol}
\begin{eqnarray}
\label{PsiSol}
    &&\Psi[C_1, C_2, t] = \nonumber\\
    &&\quad\VEV{\exp{ \I \Im \int_0^1 d \xi\Phi(\xi)}}_{\mathcal E};\\
    && \VEV{\mathcal F}_{\mathcal E} =\frac{\displaystyle\sum\displaylimits_{p<q; (p,q)}\;\; \int\displaylimits_{\Omega \in O(3)} d \Omega\int [D \alpha]\mathcal F}{\displaystyle\sum\displaylimits_{p<q; (p,q)}|O(3)|\int [D\alpha]} ;\\
    && \Phi(\xi) = \frac{e^{\imath\alpha(\xi)}\left(\mathcal C_1'(\xi|\Omega) \pm \sqrt{\Pr}\,\mathcal C_2'(\xi|\Omega)\right)}{2 \sin(\pi p/q) \sqrt{2\nu t}};\nonumber\\
    && \mathcal C_{1,2}(\theta|\Omega) =  \vect C_{1,2}(\theta) \cdot \hat{\Omega} \cdot\{\imath,1,0\};
\end{eqnarray}
\end{subequations}
Here $[D \alpha]$ is the standard path integral measure
\begin{eqnarray}
    \label{gaussMeasure}
  &&\int [D \alpha]  =  \int D \alpha(\xi) \nonumber\\
  &&\quad\exp{- \int_0^1 d \xi \frac{(\alpha')^2}{2 N \beta^2}};
\end{eqnarray}
The sum over the states of the Euler ensemble has since been computed exactly, without this continuum approximation \cite{MigdalQPTF2026}; we use that result below.
\section{Two Euler ensembles and the quantum pendulum}
Before using the exact sum over the states of the Euler ensemble \cite{MigdalQPTF2026}, let us see how each of the two loops enters the solution \eqref{solution}.

If the magnetic loop $C_2$ shrinks to a point, the polygon $\vect g_k$ drops out, and \eqref{solution} becomes the loop functional of the hydrodynamic decaying turbulence \cite{migdal2023exact, migdal2024quantum}. This is as it must be: without the magnetic force the velocity field satisfies the Navier--Stokes equation, whatever the magnetic field is.

If the hydrodynamic loop $C_1$ shrinks to a point, the polygon $\vect f_k$ drops out. The polygon $\vect g_k$ is the polygon $\vect f_k$, multiplied by $\pm\sqrt{\Pr}$. The sign is a turn by $\pi$ in the plane of the polygon, which is absorbed by the global rotation $\hat \Omega$ (the constant normal part of $\vect f_k$ does not contribute to a closed loop), and $\sqrt{\Pr}/\sqrt{2 \nu t} = 1/\sqrt{2 \eta t}$. Therefore
\begin{eqnarray}\label{magmarginal}
    &&\Psi[\,\cdot\,, C_2, t] = \nonumber\\
    &&\quad\lim_{N \to \8}\VEV{\exp{\displaystyle\frac{\I\sum_{k=1}^N \Delta \vect C_2(\theta_k) \cdot \vect f_k }{\sqrt{2\eta t}} }}_{\mathcal E}
\end{eqnarray}
This is the same Euler ensemble, with the magnetic viscosity $\eta$ in place of $\nu$.

Thus, the solution describes two Euler ensembles, one for each circulation, with the viscosities $\nu$ and $\eta$. They are not independent: the two polygons are driven by the same random walk $\sigma_l$ in the same plane, and they differ only by the factor $\pm\sqrt{\Pr}$. This synchronization affects only the joint statistics of the two circulations.

The equations leave one freedom in this synchronization: the relative parametrization of the two loops. The translation operators of the loop equation connect the points of the two loops with the same parameter $\theta$. Had we connected the point $\theta_k$ of one loop with the point $\theta_{k+s}$ of another one, the magnetic polygon would be shifted along the hydrodynamic one, $\vect g_k = \pm\sqrt{\Pr}\, \vect f_{k+s}$. This shift also affects only the joint statistics. Each of the two functionals above involves one polygon, and, as we shall see, the correlation functions of the two fields do not depend on the relative parametrization of the two loops.

Everything known about the hydrodynamic Euler ensemble applies to each of the two:
\begin{itemize}
    \item \textbf{The states.} Every admissible state of the ensemble counts once. The count is dominated by the even-$N$ polygons at zero winding, $\sum_l \sigma_l =0$, where every turning angle $\beta = 2\pi p/q$ is admissible with every balanced sequence of spins.
    \item \textbf{The fermionic trace.} The sum over the spins $\sigma_l = \pm 1$ with equal weights is a trace over fermions on the parametric circle. In the turbulent limit $N \to \infty$ at fixed turbulent viscosity $\tilde \nu = \nu N^2$, the relevant polygons are the zig-zags near the turning angle $\beta = \pi$, and the transfer operator of this trace becomes the evolution operator of a quantum pendulum.
    \item \textbf{The conditional state.} The correlation of two vorticities is the expectation conditioned on the hairpin loop of Fig.~\ref{fig::VPsi}. It is the Gibbs average over two pendulums, one for each arc of the hairpin, normalized by the partition function of the hairpin. The result is the conditional source $C(h) = \widehat G(h)/\widehat Z(h)$, a ratio of two quadratures of the explicit resolvents of the pendulums (Floquet solutions of the Mathieu equation), as a function of the separation $h$ projected on the plane of the polygon, in units of $\sqrt 2 L$, with $L = \sqrt{\tilde \nu t}$.
\end{itemize}

Computing the Wilson loop for a specific loop, say, the circle, is an interesting problem, but there is a simpler quantity.
In the next section, we are considering important calculable cases of the $\VEV{\omega\omega}, \VEV{b b}, \VEV{\omega b}$ correlation functions, where the full solution in quadratures is available.

The hydro correlation function has been directly observed in grid turbulence experiments \cite{GridTurbulence_1966} more than half a century ago and is being studied in modern large-scale real and numerical experiments\cite{SreeniDecaying, GregXi2}\footnote{The data of the decaying turbulence experiment: \url{https://drive.google.com/drive/folders/1DkHOOxhbsT0prVj65wJPnHOeb_EXc7N6?usp=drive_link} (accessed 18 November 2024).}. 

In MHD, reliable experiments are yet to be performed, as well as reliable DNS with large Reynolds numbers.
\section{Energy spectra of the two ensembles}

The simplest, and most basic observable in MHD is the covariance matrix of the vorticity/magnetic field related to the energy spectrum:
\begin{eqnarray}\label{cormatrix}
    \hat S(|\vect k|,t) = \begin{pmatrix}
        \VEV{\vect \omega \cdot \vect \omega(\vect k)} &  \VEV{\vect \omega \cdot \vect b(\vect k)}\\
         \VEV{\vect b \cdot \vect \omega(\vect k)} &  \VEV{\vect b \cdot  \vect b(\vect k)}
    \end{pmatrix}
\end{eqnarray}
\pctPDF{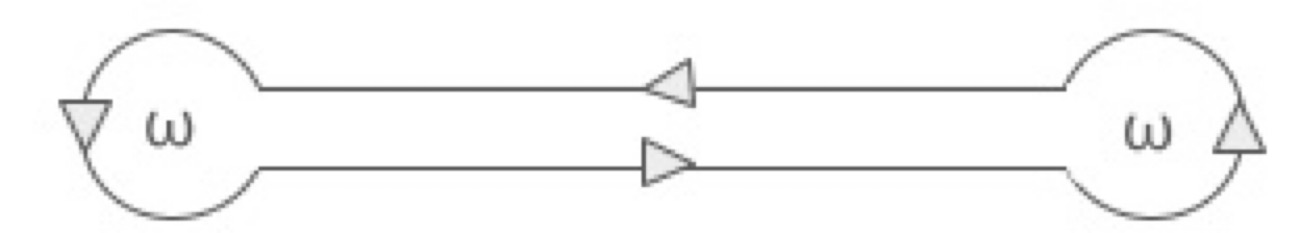}{This loop generates the vorticity/magnetic field correlation functions}
The ``hairpin'' loop shape is required to generate these correlation functions by variations in the areas of these infinitesimal circles.

Most of the computations of the correlation functions repeat those of
the hydrodynamic papers \cite{migdal2024quantum, MigdalQPTF2026} with the
replacement of $\vect f_k \Rightarrow \vect g_k$ when applied to the
magnetic field instead of vorticity.

As explained above, the statistical weight selects the even-$N$
ensemble at zero winding \cite{MigdalQPTF2026}; the odd-$N$ ensemble
remains an exact solution of
the loop equation, but its spectrum is not positive.

\subsection{Each field is measured by its own loop}
We define the correlation functions as in \cite{MigdalQPTF2026}: each of them is the expectation of the product of two fields conditioned on the hairpin that measures them (Bayes' formula). The vorticity is generated by the area derivative of the hydrodynamic loop $C_1$, and the magnetic field by the area derivative of the magnetic loop $C_2$. The correlation function of one field is a property of its marginal distribution, and the loop functional of this distribution is \eqref{solution} with the other loop shrunk to a point. Therefore, each field is measured by the hairpin of its own loop:
\begin{subequations}\label{condMHD}
\begin{eqnarray}
    &&\VEV{\vect\omega(\vect r_1) \cdot \vect \omega(\vect{r}_2)} = \nonumber\\
    &&\quad\frac{\VEV{\vect\omega(\vect r_1) \cdot \vect \omega(\vect{r}_2)\, e^{\I \Gamma_v/\nu}}}{\VEV{e^{\I \Gamma_v/\nu}}};\\
    &&\VEV{\vect b(\vect r_1) \cdot \vect b(\vect{r}_2)} = \nonumber\\
    &&\quad\frac{\VEV{\vect b(\vect r_1) \cdot \vect b(\vect{r}_2)\, e^{\I \Gamma_a/\eta}}}{\VEV{e^{\I \Gamma_a/\eta}}}
\end{eqnarray}
\end{subequations}
Here each circulation is taken around the hairpin, which runs from $\vect r_1$ to $\vect r_2$ and back, and the denominator is the amplitude of this hairpin without the two fields. In the Euler ensemble this amplitude is not equal to one, and it depends on the separation.

For the vorticity the absence of the magnetic loop is also required by the equations of motion. Without the magnetic force the velocity field satisfies the Navier--Stokes equation, whatever the magnetic field is, so that its correlation function must be that of the hydrodynamic turbulence. This is what the first definition gives, because the functional $\Psi[C_1,\,\cdot\,, t]$ is the hydrodynamic one. Moreover, each of the two definitions involves one loop only, so that it does not depend on the relative parametrization of the two loops, which the equations do not determine.

Both averages are finite sums over the states of the ensemble. The two area derivatives sit at two vertices $m <n$ of the polygon, the ends of the hairpin; the arc from $m$ to $n$ is mapped onto the wire on the way out, and the complementary arc on the way back. By \eqref{solution}, the phases of the hydrodynamic and the magnetic hairpin are
\begin{eqnarray}
    && \exp{\I \frac{(\vect r_2 - \vect r_1)\cdot \vect Q_{mn}[\vect f]}{\sqrt{2 \nu t}}};\\
    && \exp{\I \frac{(\vect r_2 - \vect r_1)\cdot \vect Q_{mn}[\vect g]}{\sqrt{2 \nu t}}};\\
    && \vect Q_{mn}[\vect f] = \frac{1}{l}\sum_{j=m}^{n-1}\vect f_j - \frac{1}{N-l} \sum_{j=n}^{m+N-1}\vect f_j;\\
    && l = n-m
\end{eqnarray}
The constant imaginary normal component of $\vect f_k$ cancels in this difference. By \eqref{aligned},
\begin{eqnarray}
   && \vect Q_{mn}[\vect g] = \pm\sqrt{\Pr}\, \vect Q_{mn}[\vect f]
\end{eqnarray}
The magnetic hairpin is the hydrodynamic one, rescaled by $\sqrt{\Pr}$.

The phase depends on the separation only through its projection $\vect r_\parallel$ on the plane of the polygon (in the strong-field problem this plane was normal to the field, and the same projection was denoted $\vect r_\perp$). The rotations in this plane, a part of the global rotation $\hat \Omega$, transform the two phases into the Bessel functions
\begin{eqnarray}
    && J_0\left(h N |Q_{mn}|\right),\quad J_0\left(\sqrt{\Pr}\, h N |Q_{mn}|\right);\\
    && h = \frac{|\vect r_\parallel|}{\sqrt 2 L};\quad L = \sqrt{\tilde \nu t};\quad \tilde \nu = \nu N^2
\end{eqnarray}
where $Q_{mn} = \vect Q_{mn}[\vect f]$. The Prandtl number enters the magnetic hairpin through the replacement of $\nu$ by $\eta$:
\begin{eqnarray}\label{Lb}
    && \sqrt{\Pr}\, h = \frac{|\vect r_\parallel|}{\sqrt 2 L_b};\quad L_b = \sqrt{\tilde \eta t} = \frac{L}{\sqrt{\Pr}};\\
    && \tilde \eta = \eta N^2 = \frac{\tilde \nu}{\Pr}
\end{eqnarray}
The number $N$ of the vertices is the same for the two polygons.

The fields at the ends of the hairpin are represented by the operators $\nu \hat \omega_1 = \I \nu\, \vect Q_1 \times \Delta \vect Q_1$ and $\eta \hat \omega_2 = \I \eta\, \vect Q_2 \times \Delta \vect Q_2$ at the two vertices. The weights $\Phi_{mn}$ of two such operators with their contact terms \cite{MigdalQPTF2026} are built of the vector products of the vectors of the polygon counted from its center. They do not change when the polygon is shifted by a constant vector or turned in its plane, and they are multiplied by $\Pr^2$ when it is rescaled by $\sqrt{\Pr}$. Together with the factor $(\eta/\nu)^2$, this leaves the weights of the hydrodynamic problem: the product of two magnetic fields at two vertices of the polygon $\vect g_k$ has the same weight as the product of two vorticities at the same vertices of $\vect f_k$. At a fixed projected separation, the numerator and the denominator of each of the two ratios \eqref{condMHD} become, up to the factor $1/(4 t^2)$ that the two operators give to the numerator,
\begin{eqnarray}
    && W_N\left(h\right),\; D_N\left(h\right) \text{ for } \VEV{\vect\omega \cdot \vect \omega};\\
    && W_N\left(\sqrt{\Pr}\, h\right),\; D_N\left(\sqrt{\Pr}\, h\right) \text{ for } \VEV{\vect b \cdot \vect b};\\
    && W_N(h) = \left\langle\sum_{m<n} \Phi_{mn} J_0(h N |Q_{mn}|) + \text{c.t.}\right\rangle_{\mathcal E}\\
    && D_N(h) = \left\langle\sum_{m<n} J_0(h N |Q_{mn}|) \right\rangle_{\mathcal E}
\end{eqnarray}
where c.t. stands for the contact term at coinciding vertices. These are the source function and the hairpin amplitude of the hydrodynamic Euler ensemble \cite{MigdalQPTF2026}. Each of them is a double sum over the positions $m, n$ of the two ends of the hairpin on the polygon, where the two fields are placed. In the continuum notation this is a double integral over the parametric circle, with the vorticity measure $\hat \omega(d \theta) = \I\, \vect Q \times d \vect Q$ and the measures of the contact terms \cite{MigdalQPTF2026}, where $\vect Q$ is counted from the center of the polygon. These measures involve only the discontinuities of the momentum loop and the fractions of its vertices on the arcs between the two points; therefore, the sums are invariant under the reparametrization of the loop.

In the turbulent limit $N \to \infty$ at fixed $\tilde \nu = \nu N^2$, both sums are dominated by the zig-zag polygons, where the ensemble reduces to two quantum pendulums \cite{MigdalQPTF2026}: the numerator tends to $\sqrt N\, \widehat G(h)$ and the amplitude of the hairpin, with the phases at its two ends removed (they belong to the observable), to $N^{\sfrac{3}{2}} \widehat Z(h)$. At a fixed projected separation the two correlations are given, up to the same factor, by the conditional source in units of $1/N$:
\begin{eqnarray}
    && \frac{1}{N}\, C\left(h\right) \text{ and } \frac{1}{N}\, C\left(\sqrt{\Pr}\, h\right);\\
    && C(h) = \frac{\widehat G(h)}{\widehat Z(h)};\quad C(0) = \frac{1}{8}
\end{eqnarray}

It remains to average over the orientation of the plane of the polygon, that is, over the angle $\vartheta$ between the separation and the normal to this plane, $|\vect r_\parallel| = r \sin \vartheta$. This can be done in two orders \cite{MigdalQPTF2026}. In the \emph{quenched} order the ratio is averaged, which gives $c$; in the \emph{annealed} order the numerator and the denominator are averaged separately, which gives $c_A$:
\begin{eqnarray}
    && c(R) = \int_0^{\sfrac{\pi}{2}} d \vartheta \sin \vartheta\, C\left(\tfrac{R \sin \vartheta}{\sqrt 2}\right);\\
    && c_A(R) = \frac{\int_0^{\sfrac{\pi}{2}} d \vartheta \sin \vartheta\, \widehat G\left(\frac{R \sin \vartheta}{\sqrt 2}\right)}{\int_0^{\sfrac{\pi}{2}} d \vartheta \sin \vartheta\, \widehat Z\left(\frac{R \sin \vartheta}{\sqrt 2}\right)}
\end{eqnarray}
The factor $1/N$ becomes $(\nu/\tilde\nu)^{\oh} = (\eta/\tilde\eta)^{\oh}$ by the renormalization at fixed turbulent viscosity \cite{MigdalQPTF2026}.

\begin{figure*}[htbp]
    \centering
    \begin{subfigure}[t]{0.48\linewidth}
        \centering
        \includegraphics[width=\linewidth]{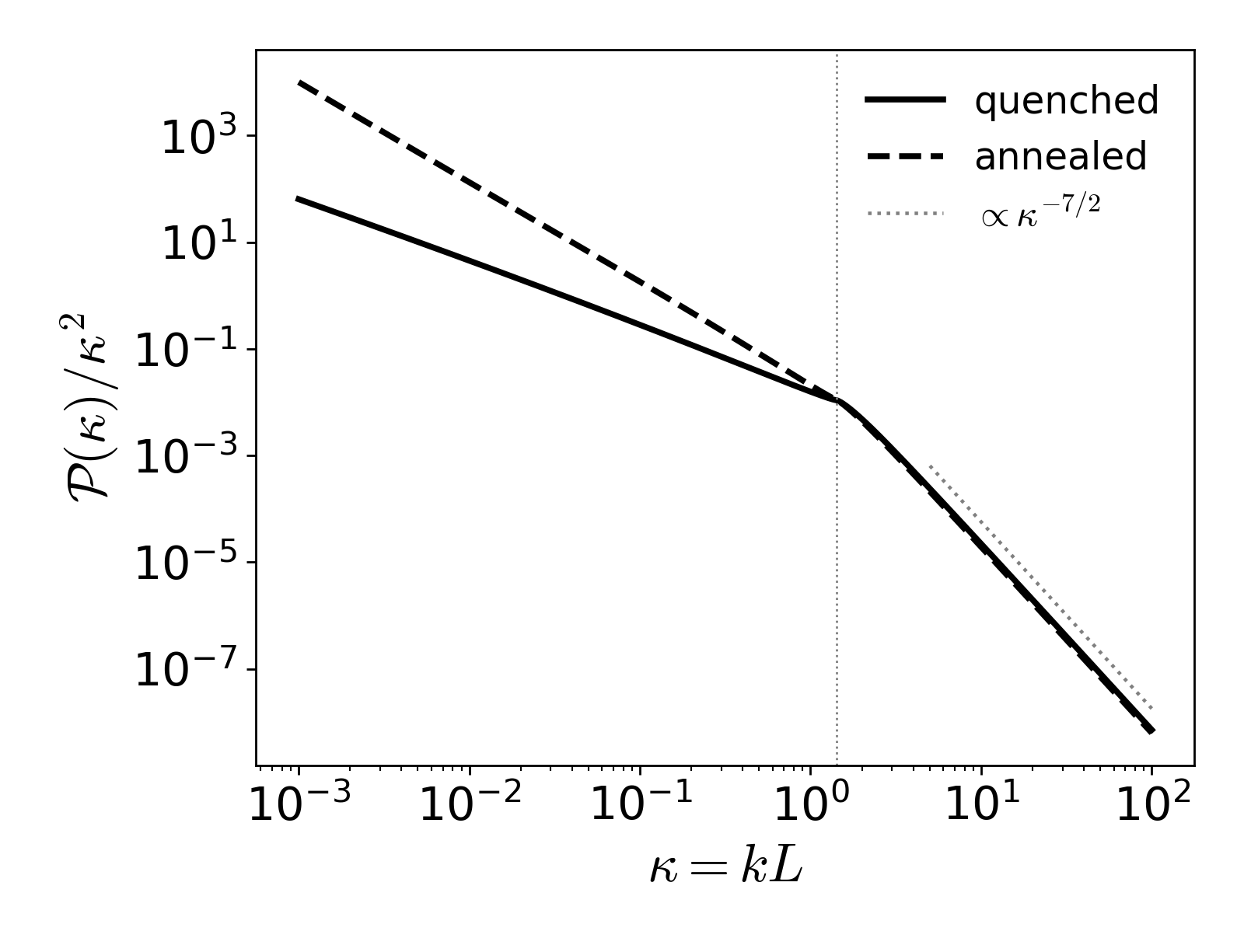}
        \caption{The kinetic spectrum $E(k,t)$ in units of
        $(\nu/\tilde\nu)^{1/2}\tilde\nu^2/(8 L)$.}
        \label{fig:KineticSpectrum}
    \end{subfigure}
    \hfill
    \begin{subfigure}[t]{0.48\linewidth}
        \centering
        \includegraphics[width=\linewidth]{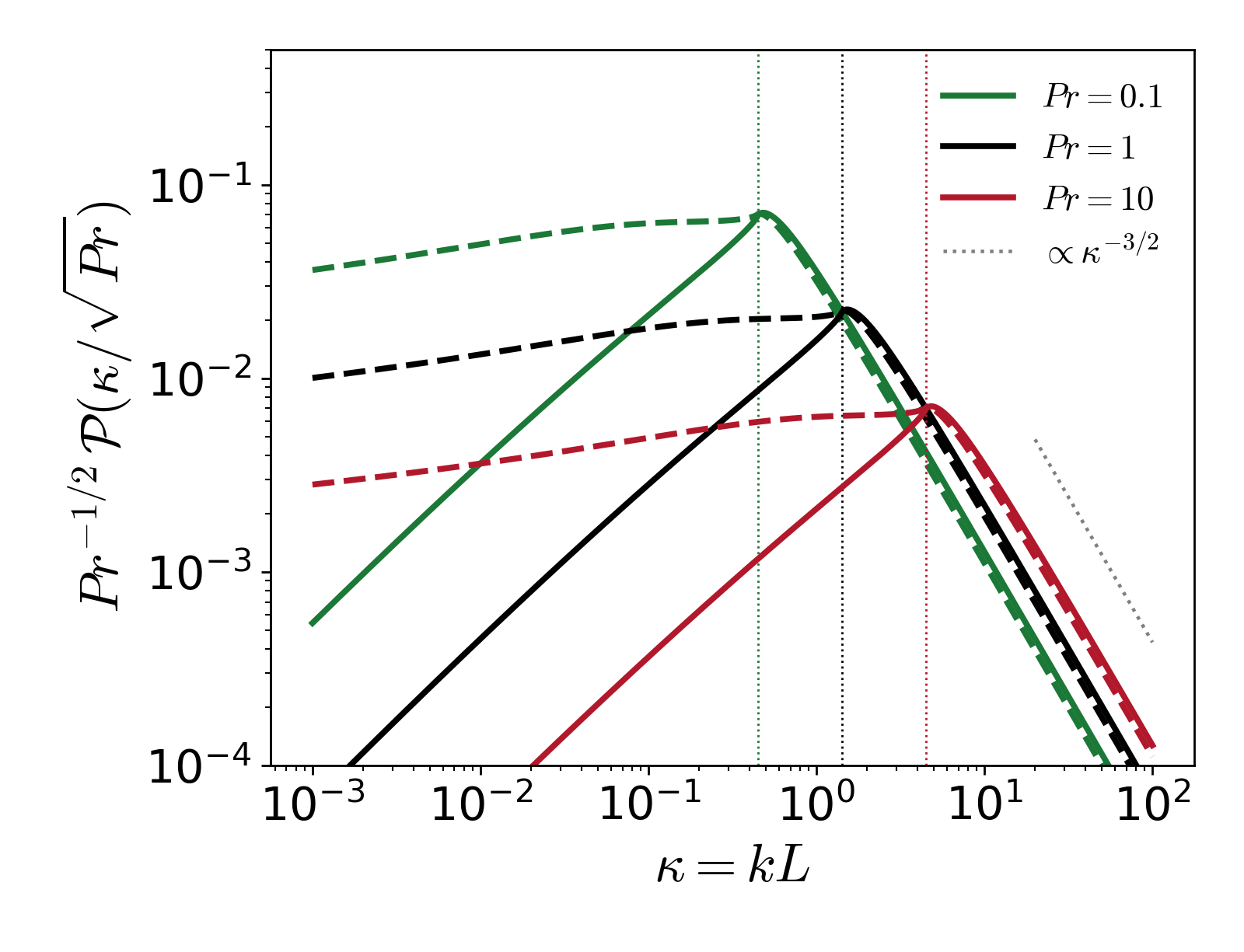}
        \caption{The magnetic spectrum $E_b(k,t)$ in units of
        $(\nu/\tilde\nu)^{1/2} L/(8 t^2)$ at $\Pr = 0.1, 1, 10$.}
        \label{fig:MagneticSpectrum}
    \end{subfigure}

    \caption{The spectra \eqref{spectrum} of the two Euler ensembles in log-log scale as
    functions of $\kappa = k L$, in the quenched (solid) and annealed (dashed) order.
    Vertical dotted lines: the thresholds $\kappa=\sqrt 2$ of the kinetic spectrum and
    $\kappa=\sqrt{2 \Pr}$ of the magnetic one. The magnetic spectrum at $\Pr = 1$ coincides with the
    enstrophy spectrum $k^2 E(k,t)$ in the same units.}
    \label{fig:TwoSpectra}
\end{figure*}

\subsection{The two correlation functions}\label{sec:twocorr}
We arrive at the correlation functions of the two fields:
\begin{subequations}\label{twocorr}
\begin{eqnarray}
    &&\VEV{\vect\omega(\vect 0) \cdot \vect \omega(\vect{r})} = \left(\frac{\nu}{\tilde\nu}\right)^{\oh}\frac{c\left(r/L\right)}{4 t^2};\\
    &&\VEV{\vect b(\vect 0) \cdot \vect b(\vect{r})} = \left(\frac{\nu}{\tilde\nu}\right)^{\oh}\frac{c\left(r/L_b\right)}{4 t^2};\\
    && L = \sqrt{\tilde \nu t};\quad L_b = \frac{L}{\sqrt{\Pr}}
\end{eqnarray}
\end{subequations}
with the same scaling function of the hydrodynamic decaying turbulence ($c$ or $c_A$, depending on the order), $c(0) = c_A(0) = \sfrac{1}{8}$. The vorticity correlation does not depend on the Prandtl number. The magnetic correlation depends on it only through its scale $L_b$: it is the vorticity correlation of a fluid with the viscosity $\eta$.

Two remarks are in order. First, here $\vect b = \vect \nabla \times \vect a$, with the vector potential normalized so that its circulation is measured in units of $\eta$, as in the loop functional. The induction equation is linear in $\vect a$; therefore the solution does not fix the amplitude of the magnetic field in other units. In Alfv\'en units the magnetic field is $\ell_0 \vect b$, with a constant length $\ell_0$, which does not enter the equations without the magnetic force, and this force vanishes on our solution; the MHD equations \eqref{NSEQ}, with the unit coefficients of the magnetic force and of the magnetic energy, are written in the units of length where $\ell_0 = 1$. In particular, the equality $\VEV{\vect b^2} = \VEV{\vect \omega^2}$, which follows from \eqref{twocorr}, refers to this normalization.

Second, the cross-correlation of the two fields. The MHD equations \eqref{NSEQ} and the initial distribution \eqref{GibbsMHD} are invariant under the reflection $\vect a \Rightarrow -\vect a$, which changes the sign of $\VEV{\omega_i(\vect r_1) b_j(\vect r_2)}$; therefore, this correlation function must vanish. In the solution \eqref{solution} the two circulations are not independent, as the two polygons are driven by the same random walk, so let us see how this comes about. The vorticity at $\vect r_1$ is generated by the area derivative of the loop $C_1$, and the magnetic field at $\vect r_2$ by the area derivative of the loop $C_2$. A variation of the loop $C_2$ does not move the points of the loop $C_1$. Therefore, the mixed derivative has no contact terms \cite{MigdalQPTF2026}: it is the product of the two operators, $\nu \hat \omega_1$ at a vertex of one polygon and $\eta \hat \omega_2$ at a vertex of another one, summed over the positions of these two vertices. Each field is measured by its own loop, and here neither loop has to pass through both points. Let us shrink the loop $C_1$ to the point $\vect r_1$, the common base point of the two loops, and keep the hairpin $C_2$, with the magnetic field at its far end. Then the phase factor does not depend on the position of the vorticity, and the sum over this position yields the total vorticity of the hydrodynamic polygon
\begin{eqnarray}\label{totalvorticity}
    \oint \hat \omega_1(d \theta) = \I \oint \vect Q_1 \times d \vect Q_1 = 0
\end{eqnarray}
This integral is invariant under the reparametrization of the loop, and it vanishes for every state of the ensemble at zero winding \cite{MigdalQPTF2026}. The same happens if we shrink the loop $C_2$ instead. Thus,
\begin{eqnarray}
    \VEV{\omega_i(\vect r_1) b_j(\vect r_2)} = 0
\end{eqnarray}
for every $N$, as the symmetry requires, and regardless of the relative parametrization of the two loops. The matrix \eqref{cormatrix} is diagonal:
\begin{eqnarray}\label{diagS}
     &&\hat S(|\vect k|,t) =
     \begin{pmatrix}
        \VEV{\vect \omega \cdot \vect \omega(\vect k)} &  0\\
         0 &  \VEV{\vect b \cdot \vect b(\vect k)}\\
    \end{pmatrix};\\
    && \VEV{\vect b \cdot \vect b(\vect k)} = \Pr^{-\sfrac{3}{2}}\left.\VEV{\vect \omega \cdot \vect \omega(\vect k')}\right|_{\vect k' = \vect k/\sqrt{\Pr}}
\end{eqnarray}

A hairpin shared by the two loops would measure the two circulations jointly. Its phase involves the sum of the two polygons, $\vect Q_{mn}[\vect f + \vect g]$, so that the result depends on the sign in \eqref{aligned} and on the relative parametrization of the two loops: with the same parametrization the separation is rescaled by $1 + \sqrt{\Pr}$ for the upper sign and by $|1 - \sqrt{\Pr}|$ for the lower one. Neither is the correlation function of the velocity field, which does not depend on the magnetic field. For the cross-correlation such a hairpin would give a term of the relative order $1/N$, which depends on the relative parametrization of the two loops and vanishes in the turbulent limit. The joint statistics of the two circulations requires a separate study.

\begin{figure*}[htbp]
    \centering
    \begin{subfigure}[t]{0.48\linewidth}
        \centering
        \includegraphics[width=\linewidth]{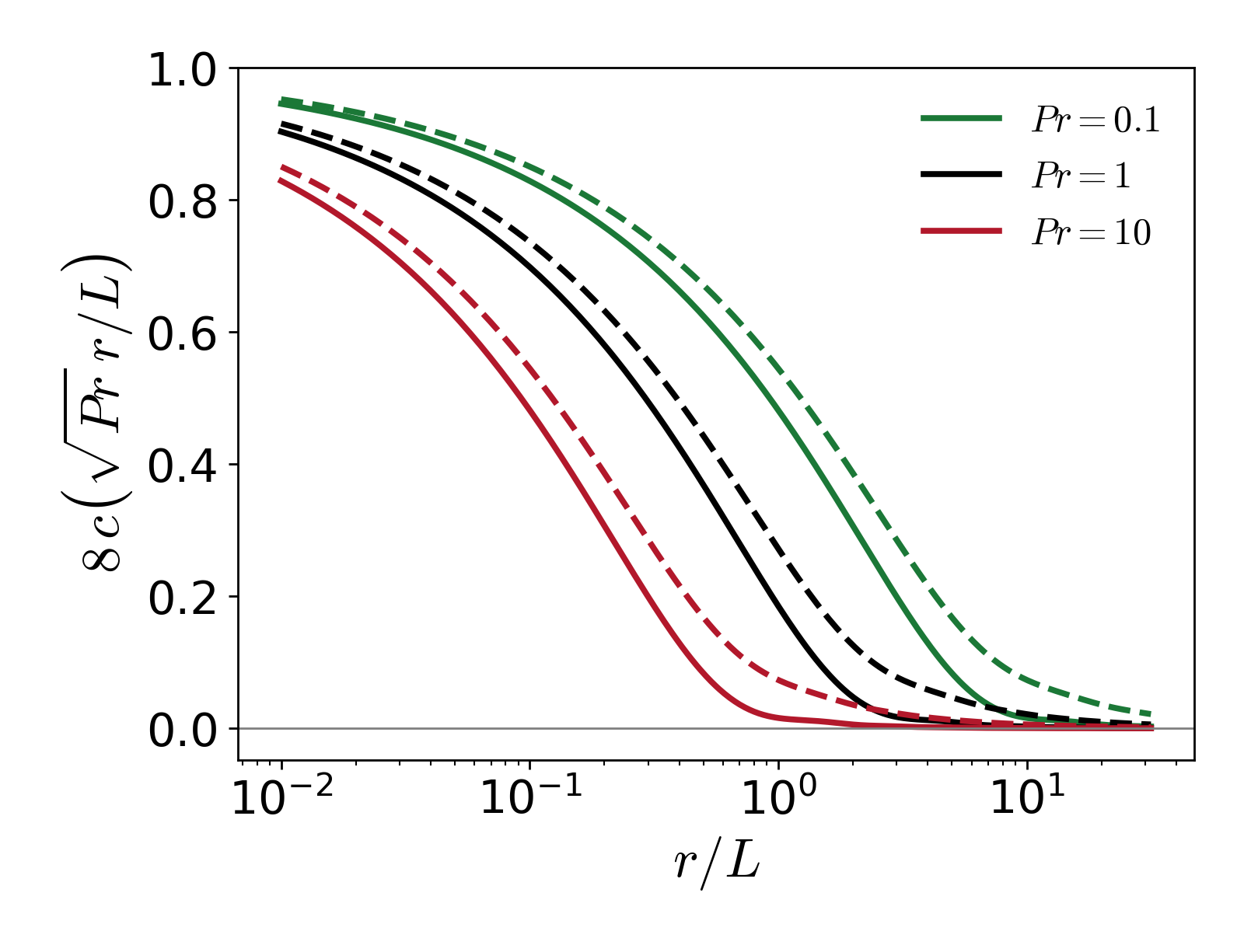}
        \caption{The correlation function \eqref{twocorr} of the magnetic field, normalized to one at zero separation,
        at $\Pr = 0.1, 1, 10$. The curve at $\Pr = 1$ is also the vorticity correlation.}
        \label{fig:CorrelationsPr}
    \end{subfigure}
    \hfill
    \begin{subfigure}[t]{0.48\linewidth}
        \centering
        \includegraphics[width=\linewidth]{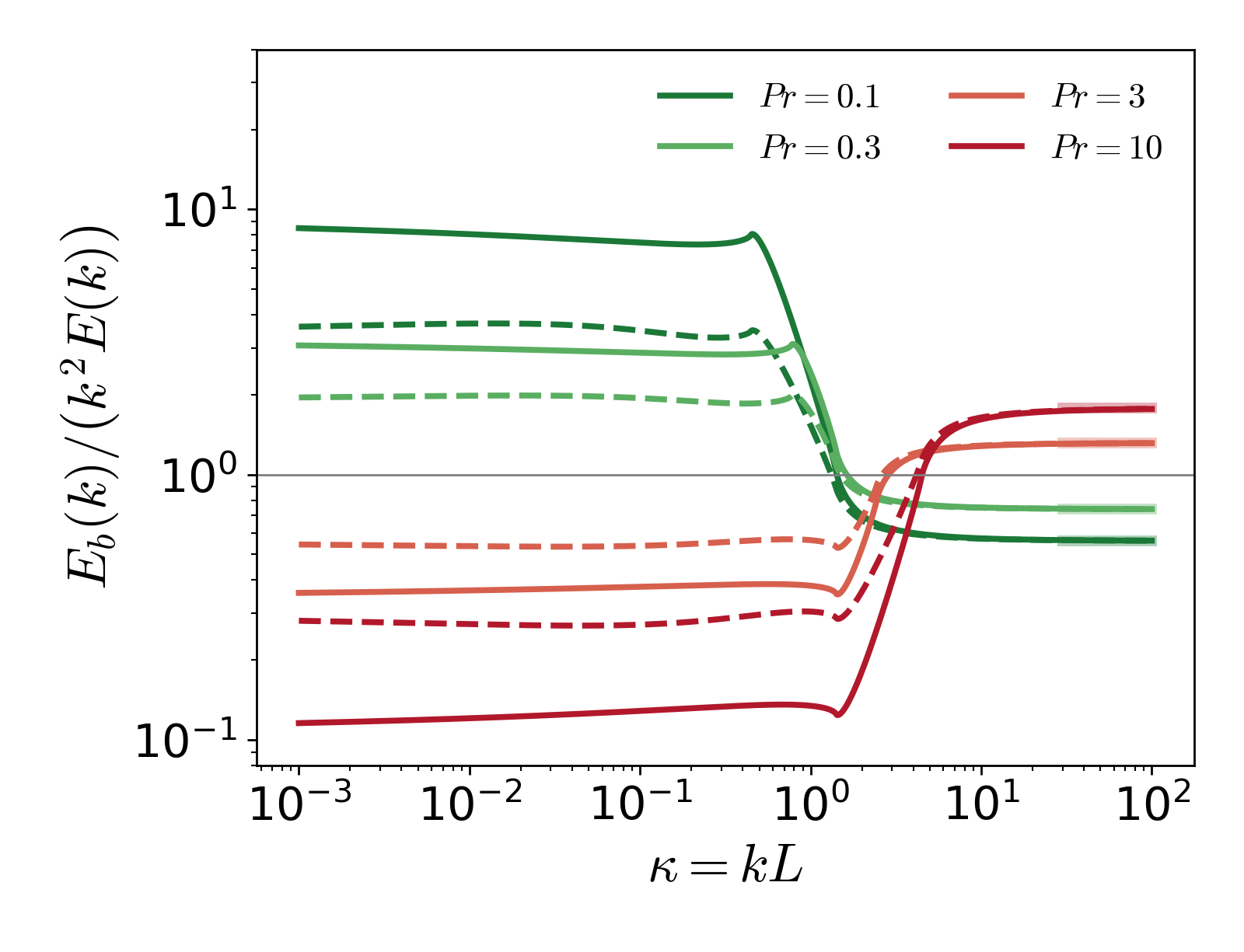}
        \caption{The ratio \eqref{ratio} of the magnetic spectrum to the enstrophy spectrum at $\Pr = 0.1, 0.3, 3, 10$;
        pale thick bars at the right edge: the limit $\Pr^{1/4}$.}
        \label{fig:SpectrumRatio}
    \end{subfigure}

    \caption{The dependence on the Prandtl number, in the quenched (solid) and annealed (dashed) order:
    (a) in coordinate space, as a function of $r/L$; (b) in Fourier space, as a function of $\kappa = k L$.}
    \label{fig:PrandtlDependence}
\end{figure*}

\subsection{The energy spectra}
The spectra follow from the relation $c(R) = \int_0^\infty d \kappa\, \mathcal P(\kappa) \sin(\kappa R)/(\kappa R)$ between the correlation and the spectral function $\mathcal P$ of the hydrodynamic problem \cite{MigdalQPTF2026}. With the normalization $\oh \VEV{\vect v^2} = \int_0^\infty d k E(k,t)$ and $\oh \VEV{\vect b^2} = \int_0^\infty d k E_b(k,t)$, we find
\begin{subequations}\label{spectrum}
\begin{eqnarray}
&&E(k,t) = \left(\frac{\nu}{\tilde\nu}\right)^{\oh}\frac{\tilde{\nu}^{2}}{8 L}\,\frac{\mathcal P\left(k L\right)}{(k L)^2};\\
&&E_b(k,t) = \left(\frac{\nu}{\tilde\nu}\right)^{\oh}\frac{L_b}{8 t^2}\,\mathcal P\left(k L_b\right)
\end{eqnarray}
\end{subequations}
The kinetic spectrum is that of the hydrodynamic decaying turbulence. The magnetic spectrum is related to the enstrophy spectrum $k^2 E(k,t)$ by the rescaling of the wavenumber,
\begin{eqnarray}\label{EbEw}
    && E_b(k,t) = \Pr^{-\oh}\, \left[k^2 E(k,t)\right]_{k \Rightarrow k/\sqrt{\Pr}}
\end{eqnarray}
and the spectrum $E_b/k^2$ of the vector potential is the kinetic spectrum with $\nu, \tilde \nu$ replaced by $\eta, \tilde \eta$. The spectrum of the magnetic energy in Alfv\'en units is $\ell_0^2 E_b$.

The universal function $\mathcal P(\kappa)$ is the same in both spectra: $\mathcal P = P$ in the quenched order and $\mathcal P = P_A$ in the annealed one.
These functions were reduced to quadratures of the resolvents of the quantum pendulum and tabulated in \cite{MigdalQPTF2026}. Both functions are positive at all computed wavenumbers, obey the sum rule $\int_0^\infty \mathcal P(\kappa) d \kappa = \sfrac{1}{8}$, and have the same threshold of the pendulums at $\kappa = \sqrt 2$, where the spectrum is continuous and the slope of $\mathcal P$ diverges. Far above the threshold both decay as $\mathcal P \propto \kappa^{-\sfrac{3}{2}}$, so that $E \propto k^{-\sfrac{7}{2}}$ and $E_b \propto k^{-\sfrac{3}{2}}$ at large $k$. For the kinetic energy, this is a faster decay than K41 model. Far below the threshold the two orders differ: $E \propto k^{-1}\log(1/k L)$ and $E_b \propto k \log(1/k L_b)$ in the quenched order, $E \propto k^{-2}/\log(1/k L)$ and $E_b \propto 1/\log(1/k L_b)$ in the annealed one. Which of the two orders a given flow follows is not decided by the theory; the comparison with the hydrodynamic DNS and wind-tunnel data is the subject of a separate paper (A.~Migdal and W.~Wang, in preparation). The magnetic spectrum has its maximum at $k L_b = 1.52$ (quenched) and $k L_b = 1.45$ (annealed).
Both spectra are shown at Fig.~\ref{fig:TwoSpectra}.

The Prandtl number enters through the ratio of the two scales, $L/L_b = \sqrt{\Pr}$ (Fig.~\ref{fig:CorrelationsPr}). The thresholds of the two spectra are at $k L = \sqrt 2$ and $k L = \sqrt{2 \Pr}$. At a given wavenumber, the ratio of the magnetic spectrum to the enstrophy spectrum is
\begin{eqnarray}\label{ratio}
    && \frac{E_b(k,t)}{k^2 E(k,t)} = \frac{\mathcal P\left(k L/\sqrt{\Pr}\right)}{\sqrt{\Pr}\,\mathcal P\left(k L\right)}
\end{eqnarray}
Above both thresholds it tends to the constant $\Pr^{\sfrac{1}{4}}$. Below both thresholds it varies slowly and approaches, logarithmically, $1/\Pr$ in the quenched order and $\Pr^{-\oh}$ in the annealed one (Fig.~\ref{fig:SpectrumRatio}). This dependence has no singularity at $\Pr = 1$: nothing happens to the two spectra at this point, where the two fields have the same correlation function.

\begin{figure*}[htbp]
    \centering
    \begin{subfigure}[t]{0.48\linewidth}
        \centering
        \includegraphics[width=\linewidth]{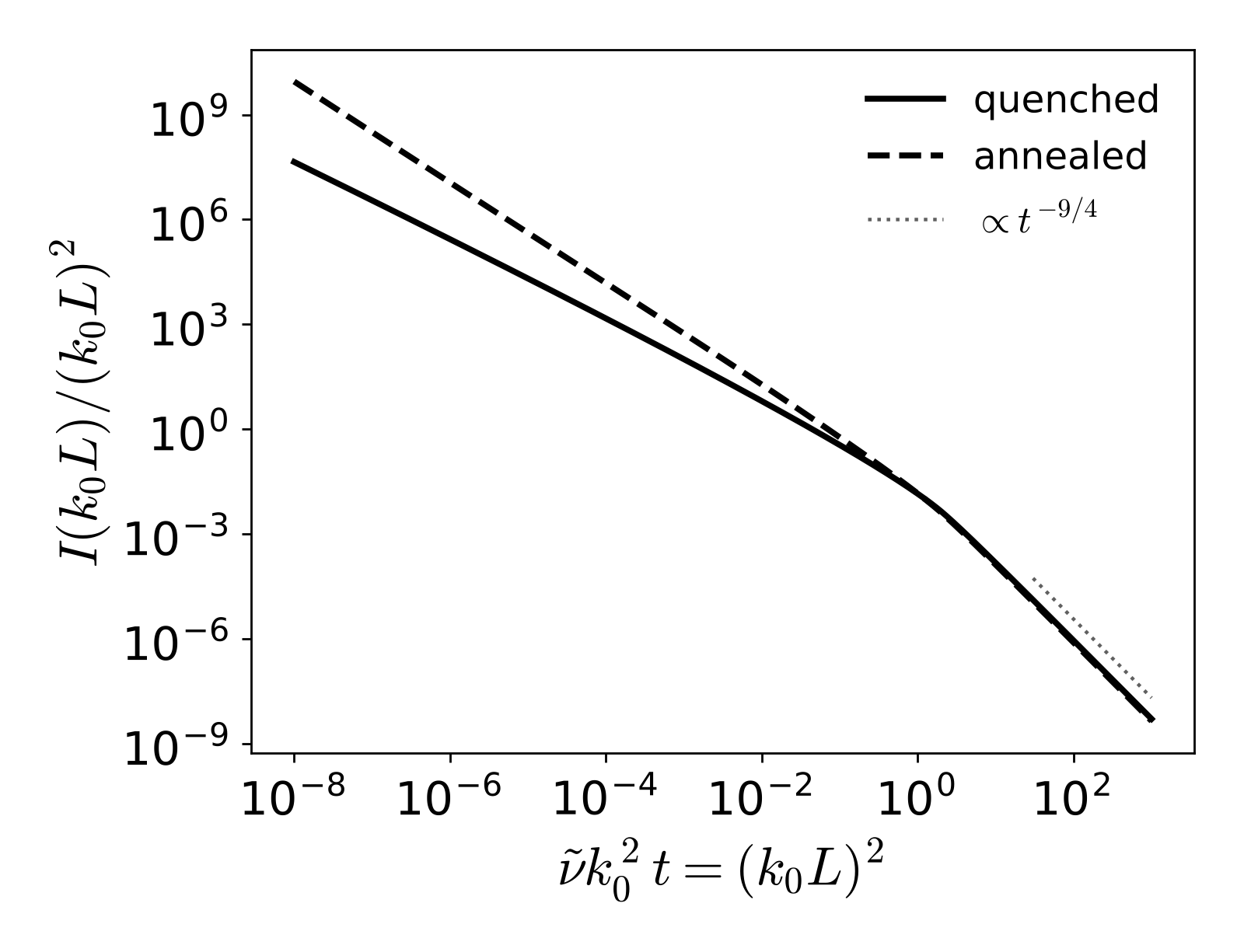}
        \caption{The kinetic energy $U(t)$ in units of $(\nu/\tilde\nu)^{1/2}\tilde\nu^2 k_0^2/8$;
        dotted line: the slope of the law $t^{-9/4}$ at $k_0 L \gg 1$.}
        \label{fig:KineticEnergy}
    \end{subfigure}
    \hfill
    \begin{subfigure}[t]{0.48\linewidth}
        \centering
        \includegraphics[width=\linewidth]{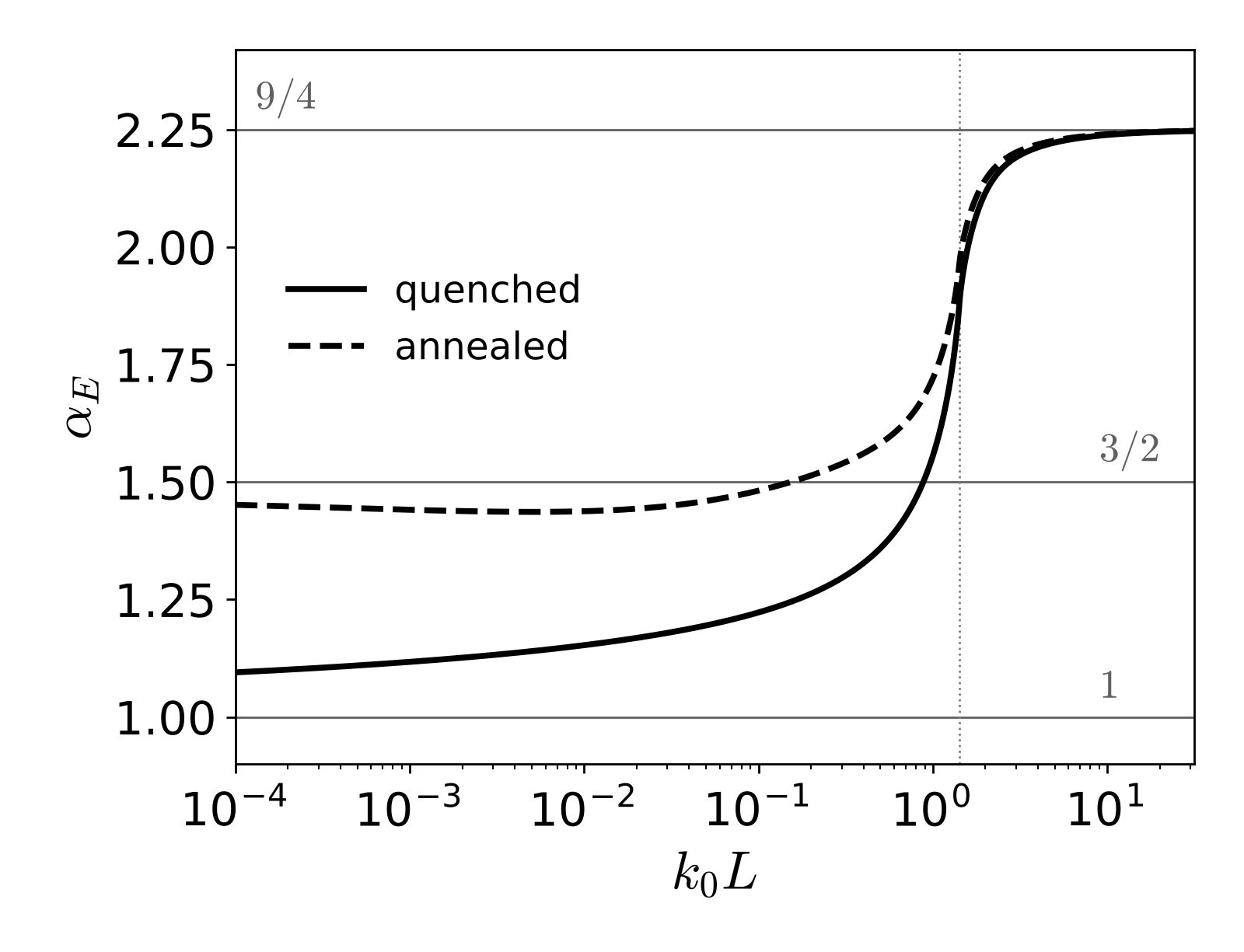}
        \caption{The decay index $\alpha_E$; horizontal lines: its limits $1$ (quenched) and $3/2$ (annealed)
        at $k_0 L \to 0$ and $9/4$ at $k_0 L \gg 1$.}
        \label{fig:KineticIndex}
    \end{subfigure}

    \caption{The decay of the kinetic energy \eqref{Uenergy} above the lowest wavenumber $k_0$, in the quenched (solid)
    and annealed (dashed) order: (a) the energy as a function of time in log-log scale; (b) its decay index as a
    function of $k_0 L = (\tilde\nu k_0^2 t)^{1/2}$. Vertical dotted line: the threshold $k_0 L = \sqrt 2$.}
    \label{fig:KineticDecay}
\end{figure*}

\begin{figure*}[htbp]
    \centering
    \begin{subfigure}[t]{0.48\linewidth}
        \centering
        \includegraphics[width=\linewidth]{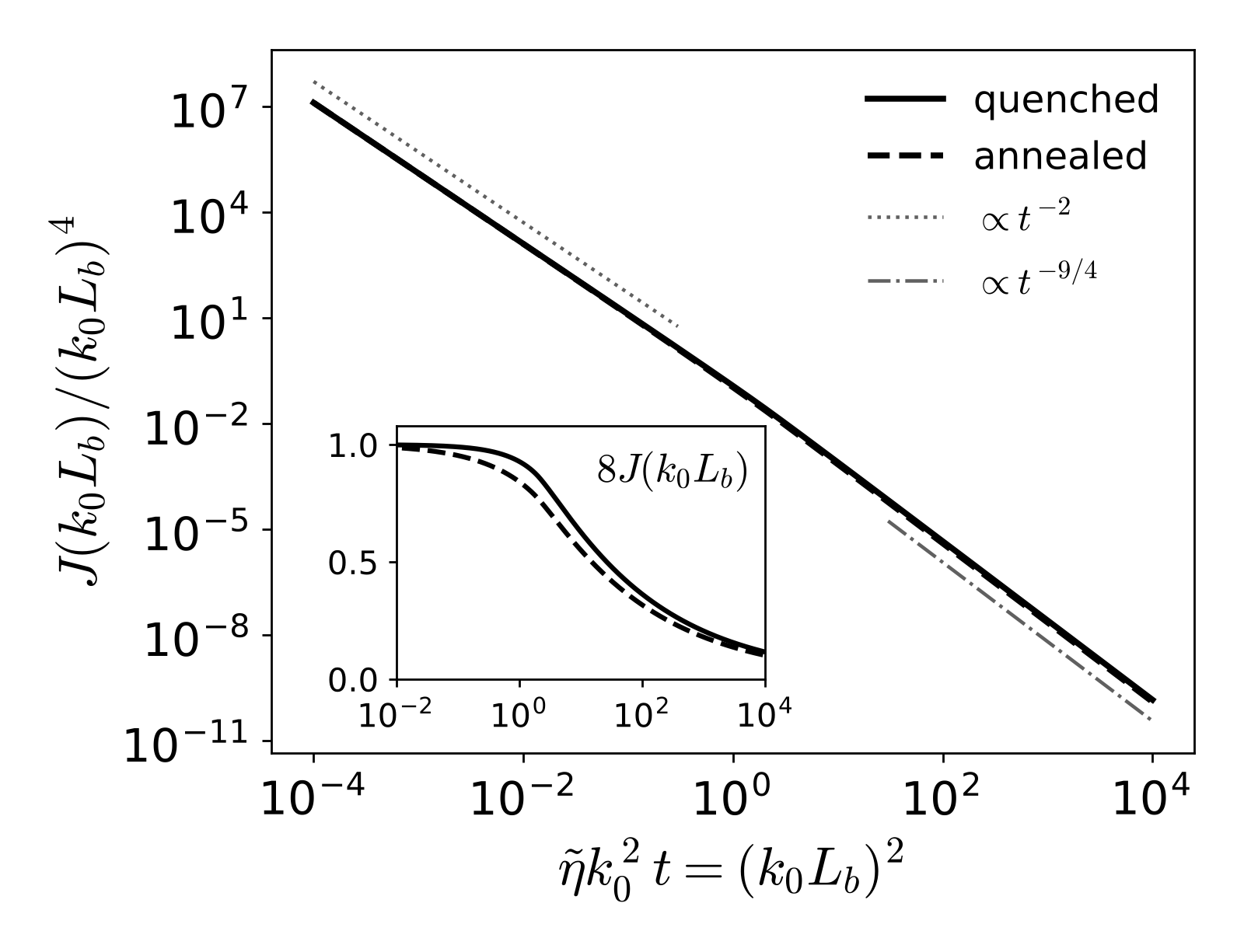}
        \caption{The magnetic energy $U_b(t)$ in units of $(\nu/\tilde\nu)^{1/2}\tilde\eta^2 k_0^4/8$;
        dotted and dash-dotted lines: the slopes of the laws $t^{-2}$ at $k_0 L_b \ll 1$ and $t^{-9/4}$ at $k_0 L_b \gg 1$.
        Inset: the same energy divided by its early law $(\nu/\tilde\nu)^{1/2}/(64 t^2)$, as a function of the same time variable.}
        \label{fig:MagneticEnergy}
    \end{subfigure}
    \hfill
    \begin{subfigure}[t]{0.48\linewidth}
        \centering
        \includegraphics[width=\linewidth]{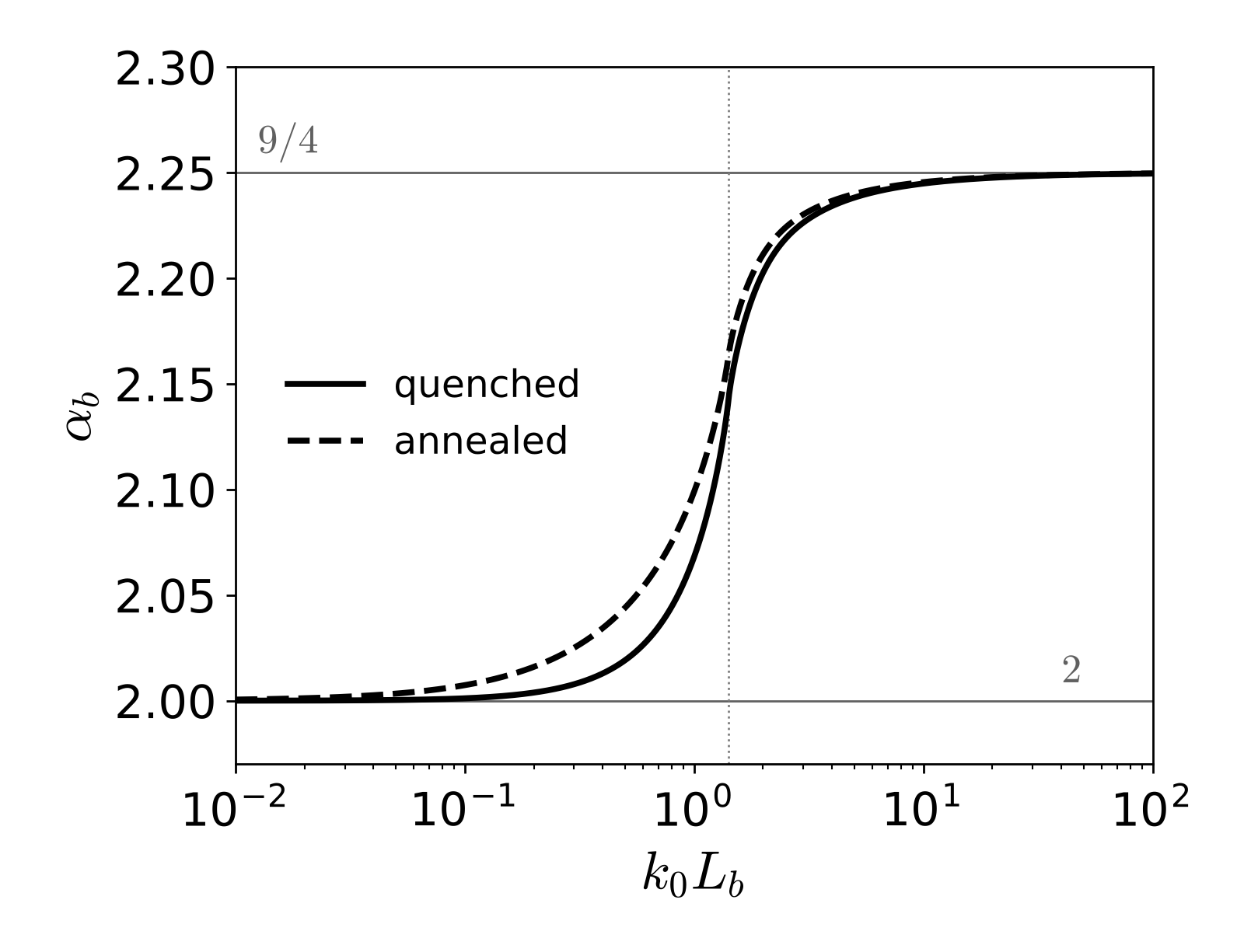}
        \caption{The decay index $\alpha_b$; horizontal lines: its limits $2$ at $k_0 L_b \to 0$ and $9/4$ at $k_0 L_b \gg 1$.}
        \label{fig:MagneticIndex}
    \end{subfigure}

    \caption{The decay of the magnetic energy \eqref{Ubenergy}, in the quenched (solid) and annealed (dashed) order:
    (a) the energy as a function of time in log-log scale; (b) its decay index as a function of
    $k_0 L_b = (\tilde\eta k_0^2 t)^{1/2}$. Vertical dotted line: the threshold $k_0 L_b = \sqrt 2$. In these variables the
    curves do not depend on the Prandtl number, which enters through $\tilde\eta = \tilde\nu/\Pr$.}
    \label{fig:MagneticDecay}
\end{figure*}

\subsection{The decay of the energy}
The kinetic energy of the pendulum spectrum diverges in the infrared (as the square of a logarithm in the quenched order and almost linearly in the annealed one), so it is defined with the lowest wavenumber of the flow, $k = k_0$, related to the inverse size of the system:
\begin{eqnarray}\label{Uenergy}
    && U(t) = \int_{k_0}^\infty d k E(k,t) \nonumber\\
    &&\quad = \left(\frac{\nu}{\tilde\nu}\right)^{\oh}\frac{\tilde{\nu}}{8 t}\, I\left(k_0 L\right);\\
    && I(x) = \int_x^\infty d \kappa \frac{\mathcal P(\kappa)}{\kappa^2};\\
    && \alpha_E(x) = -\pbyp{\log U}{\log t} = 1 + \frac{\mathcal P(x)}{2 x I(x)}
\end{eqnarray}
This energy does not decay as a power of time. Its index $\alpha_E\left(k_0 L\right)$ drifts slowly: it tends, logarithmically, to $1$ (quenched) or $\sfrac{3}{2}$ (annealed) at $k_0 L \to 0$, and to $\sfrac{9}{4}$ at $k_0 L \gg 1$. For $k_0 L$ between $10^{-4}$ and $0.3$ it lies between $1.10$ and $1.30$ in the quenched order and between $1.44$ and $1.54$ in the annealed one. It does not depend on the Prandtl number. The energy and its index are shown in Fig.~\ref{fig:KineticDecay}.

The energy of the magnetic field converges in the infrared:
\begin{eqnarray}\label{Ubenergy}
    && U_b(t) = \int_{k_0}^\infty d k E_b(k,t) \nonumber\\
    &&\quad = \left(\frac{\nu}{\tilde\nu}\right)^{\oh}\frac{J\left(k_0 L_b\right)}{8 t^2};\\
    && J(x) = \int_x^\infty d \kappa\, \mathcal P(\kappa);\quad J(0) = \frac{1}{8};\\
    && \alpha_b(x) = -\pbyp{\log U_b}{\log t} = 2 + \frac{x \mathcal P(x)}{2 J(x)}
\end{eqnarray}
As long as the magnetic scale is smaller than the size of the system, $k_0 L_b \ll 1$, this energy decays as $t^{-2}$. The index $\alpha_b$ reaches $2.07$ (quenched) and $2.10$ (annealed) at $k_0 L_b = 1$ and tends to $\sfrac{9}{4}$ at $k_0 L_b \gg 1$. The magnetic energy and its index are shown in Fig.~\ref{fig:MagneticDecay}; as functions of $k_0 L_b$ they do not depend on the Prandtl number, which enters only through the scale $L_b = L/\sqrt{\Pr}$. In Alfv\'en units the magnetic energy is $\ell_0^2 U_b$. Its ratio to the kinetic energy varies with time as $t^{\alpha_E(k_0 L) - \alpha_b(k_0 L_b)}$. At $k_0 L_b \ll 1$ this ratio is $\ell_0^2/\left(8 L^2 I(k_0 L)\right)$, and it decreases as long as $\alpha_E(k_0 L) < 2$, that is, for $k_0 L$ below $1.59$ (quenched) or $1.47$ (annealed).

\section{Possible Realization of the Large Prandtl MHD}
\begin{eqnarray}
  \frac{L_b}{L} = \Pr^{-\oh} \to 0
\end{eqnarray}
For $\Pr \gg 1$ the scale of the magnetic field is much smaller than the scale of the velocity field. The magnetic spectrum \eqref{spectrum} has its maximum at $k \approx 1.5 \sqrt{\Pr}/L$, deep inside the $k^{-\sfrac{7}{2}}$ range of the kinetic spectrum, and at larger wavenumbers it exceeds the enstrophy spectrum by the factor $\Pr^{\sfrac{1}{4}}$, in the normalization of \eqref{twocorr}. Thus, the magnetic fluctuations are concentrated at small scales, where the velocity fluctuations are already suppressed. In the solution found here these fluctuations do not grow: their energy \eqref{Ubenergy} decays as $t^{-2}$, and the kinetic spectrum is the same as without the magnetic field. The growth of a weak field by the small-scale dynamo \cite{Schekochihin2004} is not described by this solution. Systems with large Prandtl number where this regime may be looked for, at the stage when the magnetic force is small (the field is weak, or it has become force-free), include:
\begin{itemize}
    \item \textbf{Astrophysical Plasmas:} Turbulence in accretion disks, star-forming clouds, and galactic magnetic fields, where small-scale turbulence acts on a weak magnetic field.
    \item \textbf{Early Universe:} Primordial turbulence in the radiation-dominated era, potentially leaving an imprint on cosmic magnetic fields.
    \item \textbf{Quark-Gluon Plasma (QGP):} Turbulence within the QGP in high-energy heavy-ion collisions.
\end{itemize}

\section{Possible Realization of the Low Prandtl Number Regime}
\begin{eqnarray}
  \frac{L_b}{L} = \Pr^{-\oh} \to \infty
\end{eqnarray}
For $\Pr \ll 1$ the scale of the magnetic field is much larger than the scale of the velocity field. The magnetic spectrum is concentrated at $k \sim \sqrt{\Pr}/L$, below the threshold $k = \sqrt 2/L$ of the kinetic spectrum, and at $k L \gg 1$ it is suppressed relative to the enstrophy spectrum by the factor $\Pr^{\sfrac{1}{4}}$. In this sense, with the magnetic field and the vorticity normalized to the same mean square, as in \eqref{twocorr}, the magnetic fluctuations are suppressed at small scales. This regime corresponds to systems where magnetic diffusion is much faster than momentum diffusion. Systems with small Prandtl number where it may be looked for, at the stage when the magnetic force is small (the field is weak, or it has become force-free), include:

\begin{itemize}
    \item \textbf{Fusion Plasmas:} In magnetic confinement devices (tokamaks, stellarators), low-Prandtl-number MHD governs plasma turbulence, affecting transport and confinement efficiency.
    \item \textbf{Solar Wind and Magnetospheres:} The solar wind and planetary magnetospheres operate in the low-$\Pr$ regime, where strong magnetic turbulence arises due to rapid magnetic diffusion.
    \item \textbf{Earth’s Core Dynamo:} The liquid iron in Earth’s outer core has a low Prandtl number.
    \item \textbf{Liquid Metal Experiments:} Laboratory experiments using liquid sodium, gallium, or mercury to study dynamo action and MHD turbulence fall into this category.
    \item \textbf{Neutron Stars and Magnetars:} While highly magnetized, these objects may experience low-Prandtl-number turbulence in certain layers where the magnetic diffusivity is high.
\end{itemize}

The distinction between these regimes is the ratio of the two scales, $L_b/L = \Pr^{-\oh}$. Once the scale $L$ is found from the kinetic spectrum, the scale of the magnetic spectrum is predicted without free parameters. Future work should aim to compare this prediction with high-resolution DNS and experimental data.

\section{Conclusions and Perspectives}

We have developed a microscopic theory of decaying MHD turbulence, derived from first principles via the MLE equations \eqref{MLE}, \eqref{MLEdef}. The exact solution \eqref{fgsolution} describes a pair of synchronized random walks on regular star polygons: the magnetic polygon is the hydrodynamic one, rescaled by $\sqrt{\Pr}$. It exists at every Prandtl number. On this solution the operators of the velocity, the vorticity and the magnetic field are parallel at every vertex of the polygon, so that the nonlinear terms of the MHD equations vanish, and the magnetic field is force-free. Each of the two circulations is described by its own Euler ensemble, with the viscosity $\nu$ for the velocity and $\eta$ for the vector potential, and each ensemble reduces to the quantum pendulum of \cite{MigdalQPTF2026}. As a result, the vorticity and the magnetic field have the same universal correlation function, with the scales $L = \sqrt{\tilde \nu t}$ and $L_b = L/\sqrt{\Pr}$. The kinetic spectrum does not depend on the Prandtl number, and the magnetic spectrum repeats the enstrophy spectrum at the wavenumbers larger by the factor $\sqrt{\Pr}$. The synchronization of the two polygons affects only the joint statistics of the two circulations, which we leave for future work.

The decay of the energy warrants
comment in light of existing MHD DNS \cite{PhysRevE.107.055206}, which
report a Prandtl-dependent apparent exponent ($n\approx1.05$--$1.4$).
In our solution the decay index of the kinetic energy is not a
universal number. It is $\alpha_E\left(k_0 L\right)$: it depends on the
lowest wavenumber of the flow and drifts slowly with time. In the
quenched order it lies between $1.10$ and $1.30$ for
$10^{-4}<k_0 L<0.3$, and in the annealed order between $1.44$ and
$1.54$. The same is true in \emph{hydrodynamic} decay, where Rodhiya
and Sreenivasan \cite{SreeniAkash2025} found that the exponent of the
total energy is set by the infrared part of the initial spectrum, while
the bulk of the spectrum is insensitive to it; the comparison of those
data with the pendulum solution is given in a separate paper (A.~Migdal and W.~Wang, in preparation).
This index does not depend on the Prandtl number, while the magnetic
energy decays faster, as $t^{-2}$. A Prandtl-dependent apparent
exponent of the total energy can therefore come from the share of the
magnetic energy, from the lowest wavenumber, or from the magnetic
force, which vanishes only on the asymptotic solution; it is not by itself evidence against the
theory. Which of these is the case cannot be decided without the lowest
wavenumber, the scale $L$, the ratio of the two energies and the decay window of those simulations.
Because MHD couples two fields and two diffusivities, a clean
determination will require a dedicated re-analysis using the
long-time methodology of \cite{SreeniAkash2025}. We are initiating
such a study.

The reported spectra in \cite{PhysRevE.107.055206} exhibit approximate scaling with exponential cutoffs, but a clear inertial range is not resolved. Further investigation at higher resolution will be needed to verify the detailed spectral shape and assess the pendulum spectra, in particular the $k^{-\sfrac{7}{2}}$ range of the kinetic spectrum above its threshold and the scale of the magnetic one.

Future work should aim to compute the joint correlations of the velocity and the magnetic field, to study the stability of the force-free solution with respect to the magnetic force, and to determine their relevance in astrophysical and laboratory plasmas. More broadly, this theory offers a unified and tractable framework for MHD turbulence, in which scaling behavior and decay dynamics follow from an exact solution rather than phenomenological assumptions or closure models.

\section*{Conflict of Interest Statement}
The author declares that the research was conducted in the absence of any commercial or financial relationships that could be construed as a potential conflict of interest.

\section*{Author Contributions}
AM is the sole author of this work and is responsible for all of it: conceptualization, formal analysis, investigation, and writing of the manuscript.

\section*{Funding}
This research was initially supported by the Simons Foundation award ID SFI-MPS-T-MPS-00010544 in the Institute for Advanced Study in 2024; after that, it was self-funded.

\section*{Acknowledgments}
We benefited from discussions of this work with Semon Rezchikov, Alex Schekochihin, Katepalli Sreenivasan, and James Stone.

The present revision of this paper was prepared with the assistance of the generative AI model Claude Fable 5.2 (Anthropic, accessed through the Claude application in November 2026). It was used for the revision of the algebra and for the use of the quantum pendulum solution \cite{MigdalQPTF2026} in place of the earlier instanton (saddle-point) evaluation \cite{migdal2024quantum}, which is now superseded by the pendulum. The author takes full responsibility for the content of the paper.

\section*{Data Availability Statement}
The original contributions presented in this study are included in the article. The numerical tables of the universal spectral functions, from which the figures are computed, are provided with Ref.~\cite{MigdalQPTF2026}. Further inquiries can be directed to the author.

\bibliographystyle{unsrt} 
\bibliography{bibliography} 

\appendix
\section*{Computations of the Gibbs distribution}

The MHD Gibbs distribution \eqref{GibbsMHD} leads to local Gaussian fluctuations of velocity and magnetic field
\begin{eqnarray}
    && \VEV{\vect v(\vect r_1) \otimes \vect v(\vect r_2)} = T_0 F(r_{12});\\
    && \VEV{\vect b(\vect r_1) \otimes \vect b(\vect r_2)} = T_0 G(r_{12});\\
    && r_{12} = |\vect r_1 - \vect r_2|;\\
    && 4 \pi \int_0^\infty r^2 d r F(r)= 4 \pi \int_0^\infty r^2 d r G(r) = 1;
\end{eqnarray}
The support of the correlation functions $F(r), G(r)$ covers some molecular distances, much smaller than the scales of the flow $\vect v(\vect r) $ and magnetic field $\vect b(\vect r)$. 
We have to keep this microscopic scale finite at this stage, but later we can tend it to zero, and recover a continuous theory where
\begin{eqnarray}
    &&F(r_{12}) \to \delta^3(\vect r_1 - \vect r_2);
\end{eqnarray}
The loop functional can be computed for arbitrary correlation functions $F$ as an average exponential of linear Gaussian functional. 
For the hydro circulation:
\begin{eqnarray}
    &&\VEV{\exp{\frac{\I \oint d \vect C(\theta) \cdot \vect v(\vect C(\theta))}{\nu}}} = \nonumber\\
    &&\oldexp\left(-\frac{T_0}{2 \nu^2}\iint d \vect C(\theta) \cdot d \vect C(\theta')\right.\nonumber\\
    &&\quad\left.F\left(\abs{\vect C(\theta) - \vect C(\theta')}\right)\right)
\end{eqnarray}
In the local limit $\vect C(\theta') \to \vect C(\theta) + (\theta'-\theta) \vect C'(\theta)$,  this double integral reduces to the length of the loop:
\begin{eqnarray}
   &&\iint d \vect C(\theta) \cdot d \vect C(\theta')F\left(\abs{\vect C(\theta) - \vect C(\theta')}\right) \nonumber\\
   &&\to \frac{1}{r_1^2}\int d \theta |\vect C'(\theta)|;\\
   && 1/r_1^2 = \int_{-\infty}^\infty d x F(|x|) = \frac{2\int_{0}^\infty d r F(r) }{ 4 \pi \int_{0}^\infty d r r^2 F(r)}
\end{eqnarray}
For the magnetic flux, there is no analytic expression for the path integral
\begin{eqnarray}
  &&W[Q] =  \int [D C] \exp{ - \I \oint  d \theta \vect Q(\theta) \cdot \vect C'(\theta) }\nonumber\\
  &&\exp{- \oint d \theta  \oint d \theta'  \frac{T_0\vect C'(\theta)\cdot \vect C'(\theta')}{ 8 \pi \eta^2 | \vect C(\theta) - \vect C(\theta')|}}
\end{eqnarray}
The Gaussian integration over this distribution leads to an expression, well known in QED
\begin{eqnarray}
    &&\int [D a] \exp{\frac{\I}{\eta} \oint_C d \vect r \cdot \vect a(\vect r) - \int d^3 \vect r \frac{(\vect \nabla \times \vect a)^2}{2 T_0}} = \nonumber\\
    &&\exp{- \oint d \theta  \oint d \theta'  \frac{T_0\vect C'(\theta)\cdot \vect C'(\theta')}{ 8 \pi \eta^2 | \vect C(\theta) - \vect C(\theta')|}}
\end{eqnarray}
This integral logarithmically diverges at $\theta \to \theta'$ . The divergent term
\begin{eqnarray}
   &&\oint d \theta  \oint d \theta'  \frac{\vect C'(\theta)\cdot \vect C'(\theta')}{ 8 \pi | \vect C(\theta) - \vect C(\theta')|}\to \nonumber\\
   &&\quad \frac{|C|\log (|C|/r_2)}{4 \pi} ;\\
   && |C| = \oint d \theta |\vect C'(\theta)|
\end{eqnarray}
The cutoff $r_2$ corresponds to molecular distances where our approximations of the MHD break down. The coefficient is $1/(4\pi)$, because every point of the loop has neighbors on both sides; this yields $m_1 = T_0/(4 \pi \eta^2)$.
\end{document}